\documentclass[%
 reprint,
showpacs,preprintnumbers,
 amsmath,amssymb,
 aps,
prstab,
]{revtex4-1}

\usepackage{bigstrut}
\usepackage{graphicx}
\usepackage{dcolumn}
\usepackage{bm}
\usepackage{SIunits}
\usepackage{txfonts}
\usepackage[normalem]{ulem}

\usepackage[svgnames,dvipsnames,hyperref]{xcolor}
\usepackage[hyperfootnotes=true]{hyperref}
\hypersetup{colorlinks=true, linktoc=page}
\hypersetup{citecolor=RoyalBlue, linkcolor=cyan, urlcolor=red, filecolor=green}

\newcommand{\bernhard}{}

\begin{document}


\title{Testing Beam-Induced Quench Levels of LHC Superconducting Magnets in Run 1}

\author{B. Auchmann, T. Baer, M. Bednarek, G. Bellodi, C. Bracco, R. Bruce, F. Cerutti, V. Chetvertkova, \\
B. Dehning, P. P. Granieri, W. Hofle, E. B. Holzer, A. Lechner, E. Nebot Del Busto, A. Priebe, S. Redaelli, \\
B. Salvachua, M. Sapinski, R. Schmidt, N. Shetty, E. Skordis, M. Solfaroli, J. Steckert, D. Valuch, \\
A. Verweij, J. Wenninger, D. Wollmann, M. Zerlauth, \\CERN, Geneva, Switzerland\\}

\date{\today}

\begin{abstract}
In the years 2009-2013 the Large Hadron Collider (LHC) has been operated with the top beam energies of 3.5 TeV and 4 TeV per proton (from 2012) instead of the nominal \hbox{7 TeV}. The currents in the superconducting magnets were reduced accordingly. To date only seventeen beam-induced quenches have occurred; eight of them during specially designed quench tests, {\bernhard the others during injection}. There has not been a single beam-induced quench during normal collider operation with stored beam. The conditions, however, are expected to become much more challenging after the long LHC shutdown. The magnets will be operating at near nominal currents, and in the presence of high energy and high intensity beams {\bernhard with a stored energy of up to $362\;\mathrm{MJ}$ per beam}. In this paper we summarize our efforts to understand the quench levels of LHC superconducting magnets. We describe beam-loss events and dedicated experiments with beam, as well as the simulation methods used to reproduce the observable signals. The simulated energy deposition in the coils is compared to the quench levels predicted by electro-thermal models, thus allowing to validate and improve the models which are used to set {\bernhard beam-dump} thresholds on beam-loss monitors for Run 2.


\pacs{29.27.-a,41.85.Lc,29.27.Eg}

\end{abstract}

\maketitle

\section{Introduction}
During the LHC Run 1 (2009-2013) a total of 17 beam-induced quenches were observed. Most quenches occurred during dedicated experiments (quench tests) or at beam setup time. The operational quenches took place exclusively during the injection process \cite{Mariusz_LHC_quenches}. The low number of beam-induced quenches in comparison to other superconducting accelerators (HERA \cite{Kay_HERA_quenches}, \hbox{Tevatron \cite{Tevatron_quenches}}, and RHIC \cite{BIQWS_proceedings}) is explained by a better orbit stability, efficient beam-tail cleaning, sophisticated interlocks, and the low magnet currents of about half the design value. In 2015, after the Long Shutdown 1 (LS1), the LHC will be running at nominal energy and more frequent beam-induced quenches are expected. Therefore a good understanding of quench levels for various beam-loss scenarios is important, where a beam loss scenario is determined by the affected magnet, its working point, the loss duration, and the geometrical loss pattern.

The quench level is defined as the minimum local energy or power deposition that, for a given beam-loss scenario, will result in a transition from superconducting to normal-conducting state. Electro-thermal models are used to estimate the quench level. Most calculations for the LHC have been based on the phenomenological model in \cite{Note44}. {\bernhard Direct validation by measurement, however, is difficult} as spot heaters on the coil invariably alter the cooling of the strands. {\bernhard In 1977 at FNAL a magnet was installed in a beam line for test purposes. The energy deposition was measured a priori by means of a calorimeter representing the coil \cite{Edwards:1977vn}.} Here, we attempt to reproduce actual beam-loss event by means of simulation, validate the numerical model with observable monitoring signals, and take from the model the corresponding energy- or power deposition in the coils \cite{ipac14}.  The 17 beam-induced quenches in the LHC can serve to estimate quench levels in the quenching magnets. Adjacent magnets that did not quench, as well as beam-loss events that did not result in quenches at all, can serve to estimate lower bounds on quench levels. From all events, in this paper we study those that represent a relevant beam-loss scenario, that result in an energy- or power deposition in the coils sufficiently close to the assumed quench level, and that produce enough quality data for the validation of numerical models. {\bernhard The findings} are compared to electro-thermal estimates of quench levels for the respective beam-loss scenario.

In this paper we present the current status of our efforts in understanding quench levels using the example of six events, five of which were dedicated quench tests at the end of the LHC Run 1, covering a variety of beam-loss scenarios. In each case we describe the beam-loss event, explain the particle-tracking (where applicable) and the particle-shower simulations and their validation with event data, and study the consistency of electro-thermal quench-level estimates with the obtained information. In Section~\ref{QuenchLevels} we introduce terminology as well as a classification of beam-loss scenarios according to loss duration. Section~\ref{Methodology} describes the numerical analysis procedures used throughout the paper. The quench-test results are analyzed in Sections~\ref{fast} for short-duration losses, \ref{milli} for intermediate-duration losses, and \ref{steady} for steady-state losses. Section~\ref{Conclusion} summarizes the findings.  

\section{Quench levels}\label{QuenchLevels}
The quench level is a measure of the maximum amount of energy or, in the steady-state case, power that can be deposited locally in a superconducting magnet without provoking
the transition to a normal-conducting state. The quench level is a function of the local magnetic field, the operating temperature, the cooling conditions, the geometrical loss pattern, and the time distribution of the beam losses. There are three main regimes, distinguished by the duration $t$ of the beam losses:
\begin{figure}[t]
   \centering
   \includegraphics*[width=0.37\textwidth]{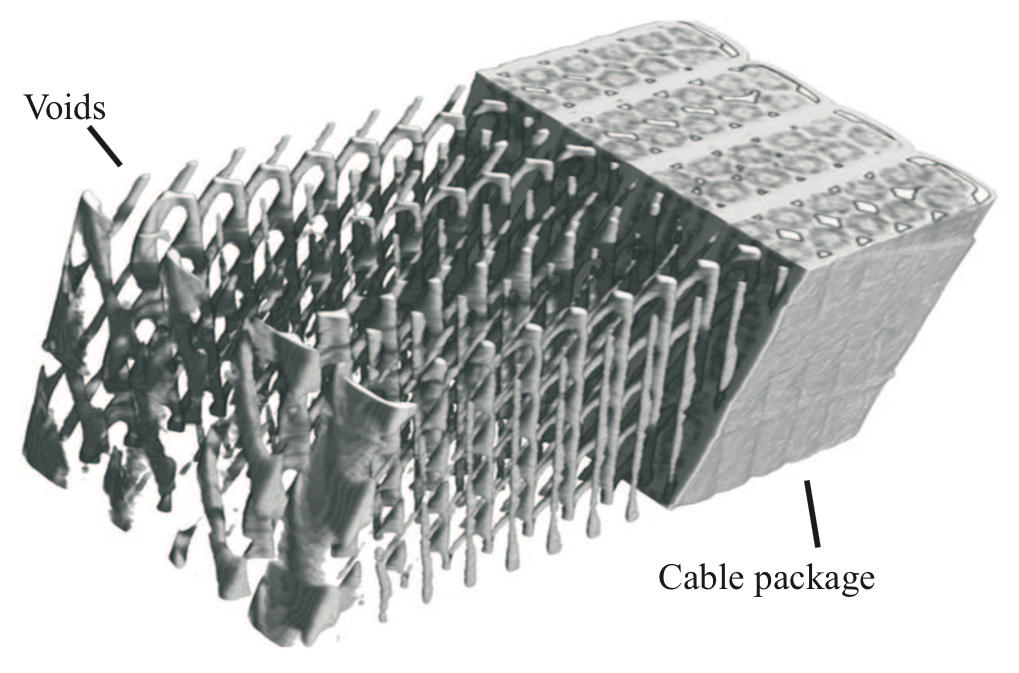}
   \caption{ \label{fig:tomography} 3-D image from neutron tomography of cable interstices in a stack of Rutherford-type cable \cite{tomography}.}
\end{figure}

{\em Short-duration} ($t <50~\mu{\rm s}$): The local quench level is determined predominantly by the volumetric heat capacity of a dry cable, with little effect of cooling to liquid helium. The quench level in this regime is quantified by the Minimum Quench Energy Density (MQED) and measured in $\rm mJ/cm^3$. In the short-duration regime, the maximum value of energy deposition across the cable cross-section is relevant. This typically coincides with the location of {\bernhard the} lowest margin to quench in the cable. The collimation quench test and the injection-study event, described in {\bernhard Sections~\ref{sec:inj} and~\ref{sec:Q6}}, respectively, probe quench levels at the sub-microsecond scale at different magnet working points.

{\em Intermediate-duration} ($ 50~\mu{\rm s}\lesssim t\lesssim 5~{\rm s}$): The liquid helium in the cable interstices and, to a lesser extent, around the insulated conductor plays a crucial role; see Fig.~\ref{fig:tomography} and Sec.~\ref{Methodology}. This is due to the efficient heat transfer to and the large heat capacity of liquid helium. In the intermediate-duration regime, the quench level is expressed by the {\bernhard above-mentioned} MQED. It depends on the actual distribution of energy deposition across the cable. The wire-scanner quench test in Section~\ref{sec:wire} and the orbit-bump quench test in Section~\ref{sec:fastADT} investigate this regime.

\begin{figure}[t]
   \centering
   \includegraphics*[width=0.5\textwidth]{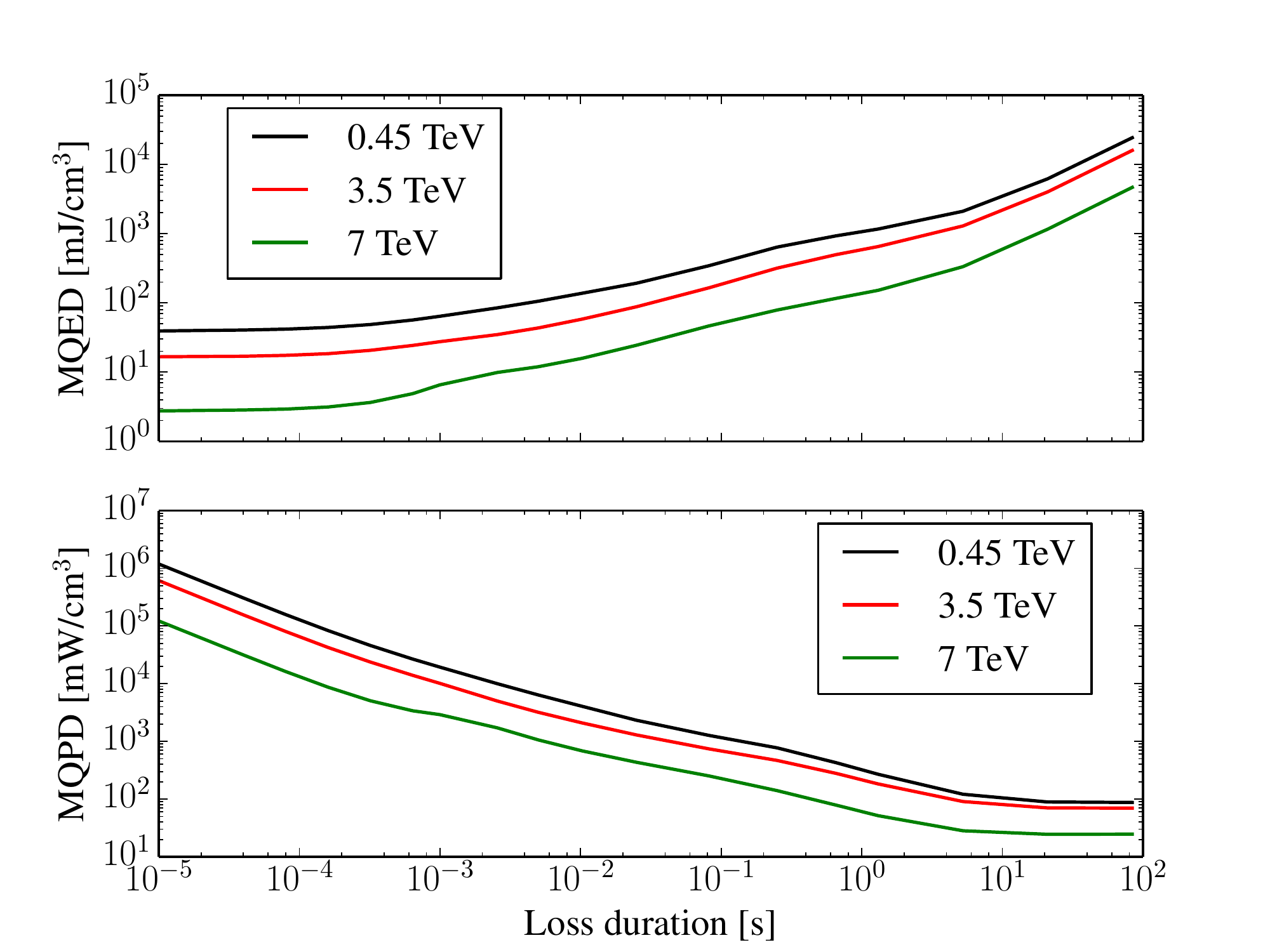}
   \caption{MQED (up) and MQPD (down) as a function of beam-loss duration for heat pulses of constant power. The quench levels are computed with QP3 \cite{Arjan_QP3} on the horizontal plane of a main-dipole magnet, for the geometrical loss pattern of \cite{LHC_Note422}, with magnet currents corresponding to injection beam energy (\hbox{450 GeV}), \hbox{3.5 TeV}, and \hbox{7 TeV}. 
   \label{F:1_qlevel}}
\end{figure}

{\em Steady-state} ($t > 5~{\rm s}$): The heat is constantly removed with a rate that is mainly determined by the heat transfer to the helium bath through the cable insulation. The quench level, in this case, is expressed as a Minimum Quench Power Density (MQPD) and measured in $\rm mW/cm^3$. MQPD is given as an average density across the cable cross-section. The collimation quench test in Section~\ref{sec:collimation} and the orbit-bump {\bernhard quench tests in Sections~\ref{sec:dynamic} and~\ref{sec:slowADT}} cover the steady-state regime.
To illustrate the dependence of the MQED and MQPD on the loss duration, Fig.~\ref{F:1_qlevel} shows simulation results of the QP3 code \cite{Arjan_QP3} for a main dipole magnet on the horizontal plane, and for the geometrical loss-distribution described in \cite{LHC_Note422}. {\bernhard It can be seen that in the short-duration regime, MQED is constant and MQPD is linear with loss duration, whereas in the steady-state regime MQPD is constant and MQED is linear. This graph was used to define the time ranges for this paper. }

\section{Methodology}\label{Methodology}
Despite the different causes of beam losses in the studied quench tests and operational events, the analysis procedures are similar in all cases. The measurement data is provided mainly by the Beam Loss Monitors (BLM), the Quench Protection System (QPS), the Beam Position Monitors (BPM), and the fast beam-current transformer (FBCT). The numerical analysis proceeds along the following steps:
\begin{figure*}[t]
   \centering
   \includegraphics*[width=0.7\textwidth]{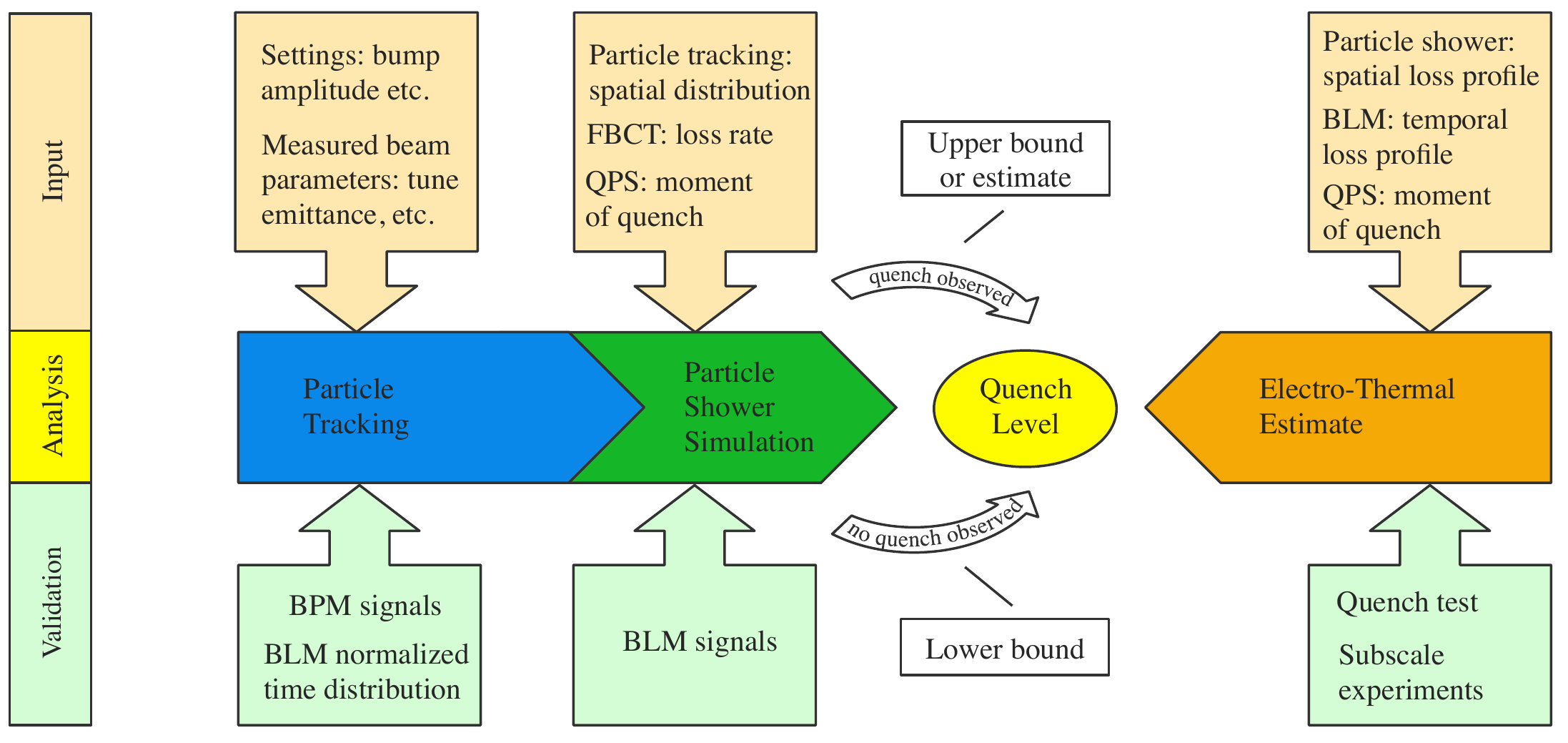}
   \caption{Overview of the analysis methodology for quench tests.  \label{fig:AnalysisOverview}}
\end{figure*}\begin{enumerate}
\item The geometric loss pattern on a suitable interface is calculated with MAD-X \cite{MADX} or SixTrack \cite{coll:sixtrack}. SixTrack, in addition to magnetic tracking, includes also a Monte-Carlo of the proton-matter interaction in the collimators, which allows multiturn tracking including out-scattering. The interface between tracking and particle-shower simulations may be the beam-screen surface, or a transverse plane, e.g., the frontal plane of a collimator. On the interface, the position- and momentum distribution of the particles serves as an input for particle-shower simulations, which may continue the tracking to the point where the particles hit dense matter. Particle-tracking with MAD-X or SixTrack may cover the moment of maximum losses, a steady-state regime, or all the beam manipulations leading up to the beam loss.
\item Particle-shower simulations with FLUKA \cite{FLUKA,FLUKAb} or Geant4 \cite{Geant4} {\bernhard are used to estimate} the energy deposition in the BLMs' active volume and inside the superconducting coil. Longitudinally, the simulation may cover a single magnet or a whole section of the accelerator. Radially the simulated geometry extends to the tunnel walls. Particle-shower simulations provide distributions of deposited energy per impacting proton. The normalisation of the simulation results is done by means of the total loss in beam intensity as measured by the FBCTs. The energy deposition in the BLMs is compared to the measured signal. Good agreement gives confidence in the simulated energy deposition in the coils. A full account of the methodology and related uncertainties of particle-shower simulations related to LHC beam operation is given in \cite{AntonFLUKA}. 
\item An electro-thermal simulation with QP3, THEA~\cite{THEA}, or ZeroDee~\cite{ZeroDee} yields quench level estimates in the most critical position of the coil. QP3 and THEA are 1-dimensional codes, taking the distribution of losses along a strand into account, whereas the averaging assumptions of ZeroDee make it suited for strand-wise computations in the short-duration regime, and cable-computations in the steady-state regimes only. Both, QP3 and THEA, provide different options to take the cooling to helium inside the cable into account. For the same assumptions, they yield identical results. Electro-thermal estimates presented in this paper are based on the heat-transfer models documented in 
\cite{bauer}. For the intermediate and steady-state regimes, the BLM signal provides the time profile of the heat pulse in the coils. The time profile is curtailed at the moment a resistive voltage is visible in the QPS data. \label{BLMQPS}The radial loss profile across the cable from FLUKA, the temporal profile of losses from the BLM signals, and the magnetic field distribution across the cable from ROXIE \cite{ROXIE} at the given magnet current are taken into account. Note that only relative, not absolute values of BLM signals and FLUKA simulations are used as input to the electro-thermal model. The influence of the radial distribution from FLUKA on the electro-thermal quench-level estimate is significant only in the intermediate-duration regime, where it may change the computed MQED by several ten percent.

\item {\bernhard If no quench occurred in the simulated event,} particle-shower simulations provide a lower bound for the quench level. If a quench {\bernhard occurred, the} {\bernhard energy deposition based on} the total number of protons lost in the event provides an upper bound. For intermediate-duration and steady-state losses, the determination of the moment of quench during the beam loss {\bernhard period} allows {\bernhard to determine the number of protons lost at the moment of quench, and, thus,} to deduce a direct estimate of the quench level. Consistency between particle-shower simulations and quench-level estimates increases the confidence in the electro-thermal model as well as in the overall understanding of the event.
\item The electro-thermal model can be used to extrapolate the quench level estimate to similar events at different magnet currents.
\end{enumerate}
Figure~\ref{fig:AnalysisOverview} illustrates the analysis process. Both the numerical simulations and the experimental data are affected by errors. We mention the most important {\bernhard ones}:

{\em Electro-thermal models} are affected by three major uncertainties:
(1) In the intermediate-duration regime, the model features multiple mechanisms of heat transfer between the strands and the helium filling the voids in the Rutherford cable (Kapitza cooling, convection cooling, nucleate and film boiling, etc.). Models and parameters vary widely in literature. For this paper we use two distinct models described in \cite[Sec.~2.2]{pbauer_PhDthesis} and \cite{ppgranieri_transient}, respectively. The differences in the results are presented as an uncertainty range in this paper.
(2) Mostly affecting the intermediate-duration regime is the amount of helium in the inter-strand voids of the Rutherford cable, as well as the area of contact between strands and helium. Tomographic imaging was used in \cite[p.~59ff.]{gerard_PhDthesis} 
and the results were in agreement with previous measurements in \cite{depond}. In \cite{ppgranieri_transient}, the more pragmatic assumption is made that half of the geometric void area is filled with helium (the other half being filled with the Kapton insulation), and that 50\% of the strand diameters are in contact with helium. 
(3) Experimental work on the heat extraction through the cable insulation in the steady-state regime has been carried out at CERN \cite{Richter:2007pi,Granieri:2010hc,ppgranieri_steadystate}. The experimental data used in the QP3 and THEA codes is described in \cite{Granieri:2010hc}. More experimental work is under way to confirm and extend this data set, in particular with regard to the efficiency of the intra-layer spacers in the LHC main magnets \cite{ppgranieri_BIQ}.

{\em Particle-shower simulations} rely on an approximation of the equipment and tunnel geometries, as well as material distributions therein. Simulations that require a model of an extended section of the accelerator cannot be modeled with the same level of detail as those that require only one or two magnets. The geometry of coil ends is not accurately modeled in FLUKA, so that energy depositions in the magnet ends are not evaluated with the same accuracy as those in the magnet straight sections. Cases that result in a pronounced peak in energy-deposition may suffer from the averaging over evaluation cells that are usually 10~cm long. Statistical errors are typically negligible. Geant4 simulations of several of the quench tests are discussed in \cite{Agnieszka_PhDthesis}; to make their models more generic, the authors assumed a constant impact angle of particles on the beam screen. Here we present work with FLUKA in which each model attempts to represent the actual events and tests as accurately as possible. More detailed information is found in \cite{AntonFLUKA}.

{\bernhard {\em Particle-tracking} uncertainties arise from an idealized description of the machine} and imperfect knowledge of initial conditions. Geometric parameters such as tolerances on the beam-screen geometry, surface roughness, and {\bernhard misalignment} affect the results. Additional uncertainties in the case of SixTrack come from the simulation of particle-matter interactions in collimator jaws.

{\em QPS data} is provided for system monitoring at 5~ms intervals. For losses in the intermediate-duration regime this sampling rate makes a precise determination of the moment of quench difficult; see point~\ref{BLMQPS} above. The synchronisation of BLM data and QPS data is affected by a similar uncertainty of 5-10~ms.

{\em BLMs} that are exposed to large energy deposition draw high currents from their power distribution line. In rare cases this may affect the reading on other BLMs on the same power-distribution line.


\section{Short-Duration Losses}\label{fast}

The most likely loss scenario in this regime are an injection failure {\bernhard or an asynchronous beam dump}. The quenches induced by fast beam losses at injection are described in \cite{LHC_Note422,Anton_fastloss}. While an injection failure will quench magnets, the collimation system and the QPS protect the magnets from damage; the BLM system is used for a posteriori diagnostics. As for asynchronous beam dumps, the most affected magnets are the quadrupoles close to the dump kickers called Q4 and Q5 (wide-aperture quadrupoles), operating at 4.5~K. A study of quench levels for the short-duration regime is used to set trigger levels for abort-gap cleaning \cite{Bernhard_ABD}. In the following, two beam-induced quenches are investigated. First, an actual event from the first commissioning with beam, and then a dedicated quench test to probe quench levels in the short-duration regime.

\subsection{Strong-Kick Quench Event}\label{sec:750urad}
\label{sec:inj}


{\bf \noindent  Experimental setup -- } Out of several beam-induced quenches which took place at injection \cite{Mariusz_LHC_quenches}, an event of $9^{\mathrm{th}}$ September 2008 is presented here. In this event, a main dipole (MB.B10R2 operating at 1.9 K) in the dispersion suppressor region, equipped with several BLMs, was quenched in an attempt to reproduce an accidental quench that occurred a few weeks earlier during an aperture scan in the arc downstream of IR 2 (Interaction Region 2). The earlier quench concerned a different magnet. The measurement consisted of injecting a pilot bunch (few 10$^9$ protons), inducing trajectory oscillations with {\bernhard various} amplitudes by means of a vertical corrector (MCBCV.9R2), and monitoring the downstream losses. The beam was then stopped at a collimator in the momentum cleaning insertion (IR3). A large vertical kick of nominally 750~$\mu$rad was applied at the corrector and 2$\times$10$^9$ protons hit directly the MB.B10R2 aperture, inducing a quench.

{\bf \noindent  Particle tracking -- }The reconstruction of the trajectory of the kicked {\bernhard beam was} done with MAD-X by matching the calculated trajectory with the actual BPM measurements. The strengths of the correctors, which were employed to correct the trajectory to the reference one, were used as matching variables (the range is $\pm$20\% of the applied strength). {\bernhard The initial beam coordinates ($x_0$, $x_0$', $y_0$ and $y_0$') were used to better fit the BPM signals, thus, accounting for possible injection errors} \cite{chiara_ipac14}. 

\begin{figure}[t]
\centering
\includegraphics*[width=85mm]{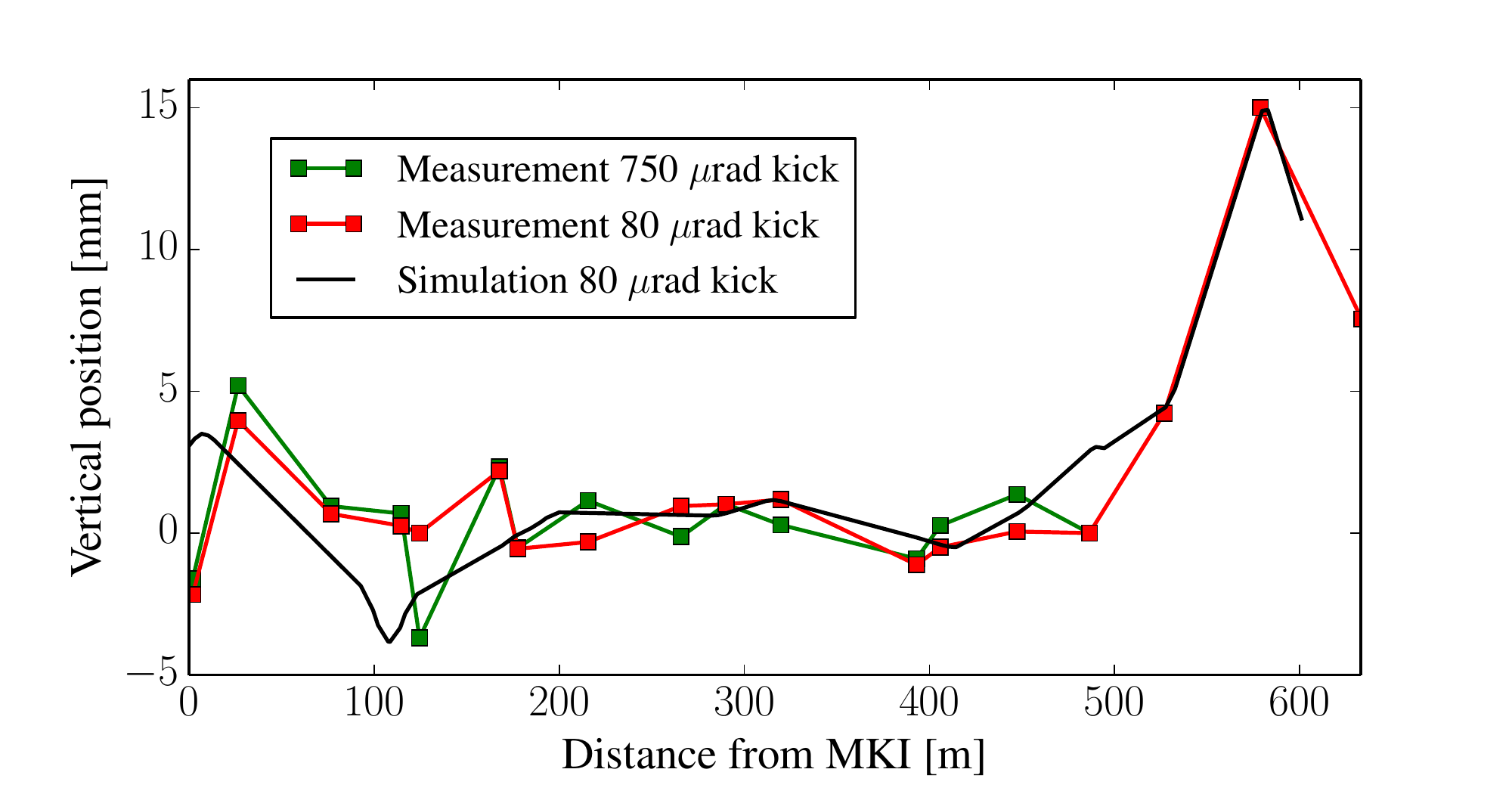}
\caption{{\bernhard Vertical trajectory of the injected beam. Black line: MAD-X simulation of injection with 80~$\mathrm{\mu rad}$ kick by MCBCV.9R2 vertical orbit corrector. Red line: BPM readings for of 80~$\mathrm{\mu rad}$ kick. Green line: BPM readings for 750~$\mathrm{\mu rad}$ kick.}}
\label{fig:large-kick_MADX_vertical}
\end{figure}

For the shot which caused the quench, {\bernhard no BPM data} was available downstream of the MCBCV.9R2 orbit corrector (green line in Fig.~\ref{fig:large-kick_MADX_vertical}, the corrector is located at 494~m from the injection kickers MKI) {\bernhard since the beam was lost due to the large kick}. Data from a previous injection, with an applied kick of 80~$\mu$rad, was used as reference for the matching and a reasonable agreement with measurements was found as shown in Fig.~\ref{fig:large-kick_MADX_vertical}. These calculations allowed to define the position of the beam ($\pm$1~mm accuracy) at the vertical corrector and the real strength of the kick (714~$\mu$rad~$\pm$10\%) given to the beam when the quench occurred. 

%

\begin{figure}[t]
\centering
\includegraphics*[width=0.5\textwidth]{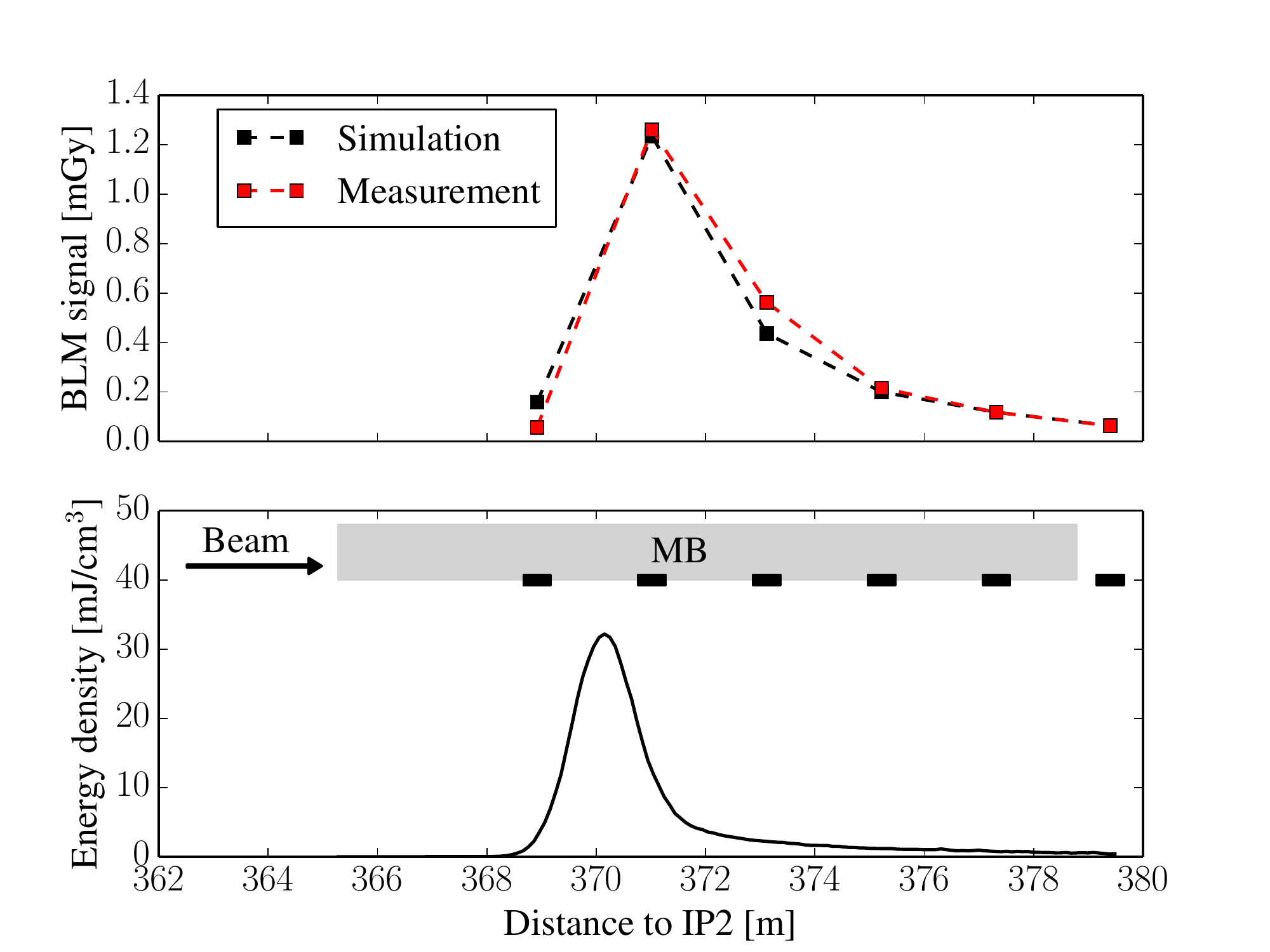}
\caption{{\bernhard Up:} Comparison of BLM signals measured during the strong-kick event and simulated with FLUKA. Beam direction is from the left to the right. {\bernhard Down: FLUKA simulated peak energy deposition in the coils, integrated over the event. Gray box indicates magnet cold mass, black boxes indicate the location of BLMs.}}
\label{fig:large-kick_BLM}
\end{figure}

{\bf \noindent  Particle-shower simulation -- }{\bernhard A transverse plane at the beginning of the MCBCV.9R2 orbit corrector acted as the interface between MAD-X and FLUKA simulations.} The coordinates {\bernhard and momenta at the interface} were used as {\bernhard starting values in the FLUKA energy deposition studies} in the dipole located 24~m downstream. Figure \ref{fig:large-kick_BLM} {\bernhard (up)} shows the measured maximum BLM signals compared to the simulated maximum BLM signals during the event. The agreement observed for all the downstream BLMs is within 20\%. In this particular event, BLM signals are very sensitive to uncertainties in the vertical coordinates at the interface between particle-tracking and particle-shower simulations. {\bernhard Figure \ref{fig:large-kick_BLM} (down) shows the maximum deposited energy density along the coil.} The energy deposition in the coil is very sensitive to the horizontal angle at the interface. 

\begin{figure}[t]
\centering
\includegraphics*[width=0.4\textwidth]{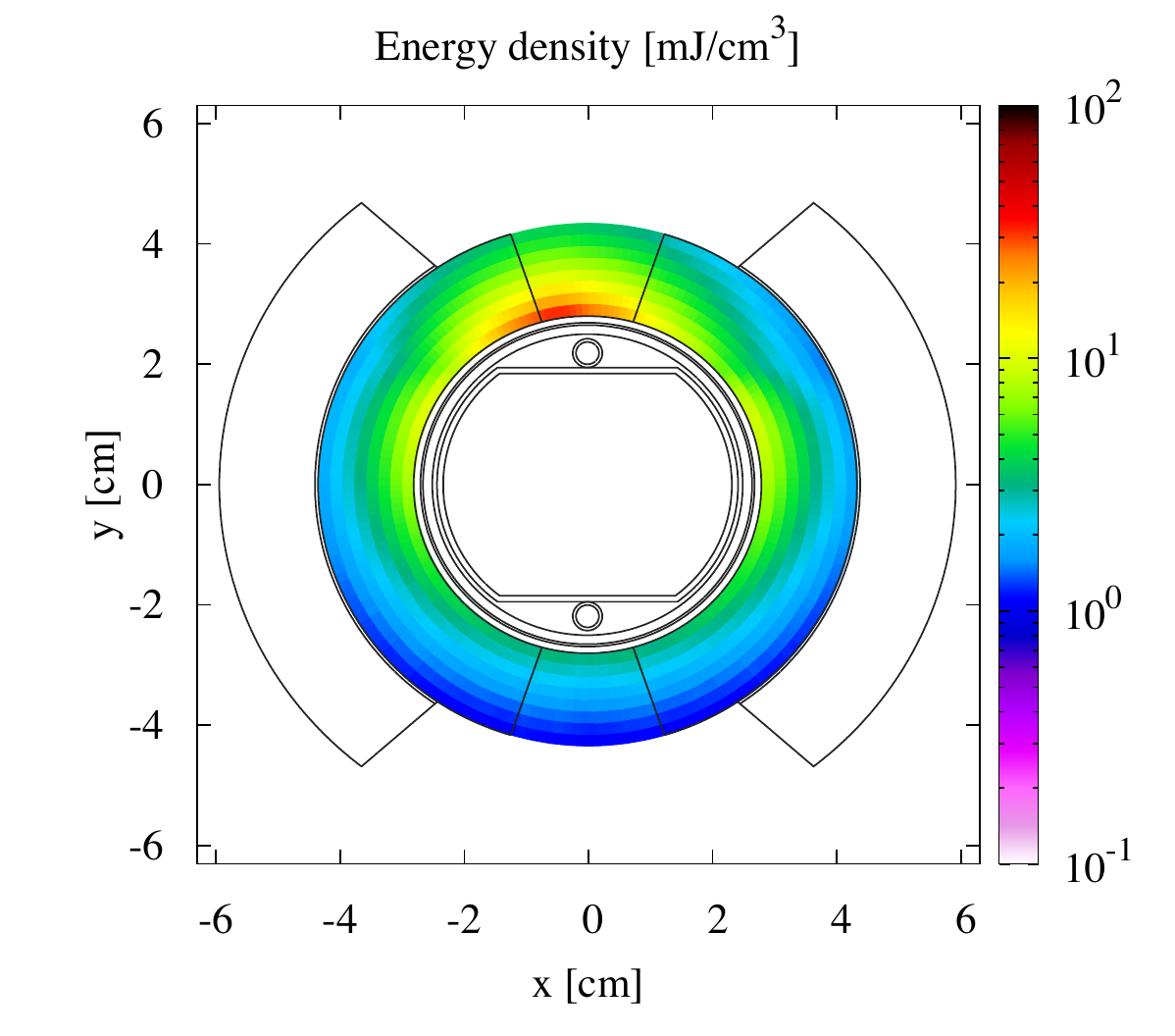}
\caption{Simulated transverse energy density distribution {\bernhard from FLUKA} for the strong-kick event in MB.B10R2 coils at the position where the maximum energy deposition occurs. Results correspond to 2$\times$10$^9$ protons impacting on the magnet beam screen. Spatial coordinates are with respect to the center of the vacuum chamber.}
\label{fig:large-kick_ED_transverse}
\end{figure}

The resulting energy density map is shown in Fig.~\ref{fig:large-kick_ED_transverse}. The maximum energy density in the coil is about 36\,mJ/cm$^3$, which occurs very close to the vertical plane. For the beam emittance a conservative estimate of 1 $\mu$m$\cdot$rad was used. If the emittance {\bernhard were} smaller, the maximum energy density in the coil {\bernhard could have been} higher. For a vertical kick of a beam on the design trajectory, the maximum of the deposited energy density would have been found in the magnet collar; due to oscillations of the beam after injection, the maximum energy density is moved into the magnet coil. This effect accounts for a large part of the discrepancy with analysis results presented in \cite[p.~37]{LHC_Note422}.



{\bf \noindent  Electro-thermal simulation -- } 
Results of FLUKA and of electro-thermal analyses are shown in Tab.~\ref{tab:LK:analyses}. The losses being instantaneous, helium cooling does not play a role in the electro-thermal model. The MQED can be calculated directly from the effective heat capacity of the strand according to NIST material data \cite{nist1,nist2}. The electro-thermal simulation codes agree with the value thus obtained. The uncertainty in the particle tracking induces an uncertainty in the FLUKA simulations, which may well account for the small discrepancy between the upper bound obtained from FLUKA and the calculated MQED.
\begin{table}[!t]
\caption{Quench-level comparison; FLUKA upper bound and the electro-thermal MQED estimate for the strong-kick event.}
\label{tab:LK:analyses}
\centering
\begin{tabular}{cc}
\hline
{ Particle Shower }& Electro-Thermal \\
Calculation & Estimate\\
\, [mJ/cm$^3$]& [mJ/cm$^3$]\\
\hline
  $\leq$36 & 38\bigstrut \\
\hline
\end{tabular}
\end{table}

{\bf \noindent Discussion -- }The simulation of beam-losses due to a kick or an orbit-bump requires an accurate model of the beam dynamics leading up to the loss. We will encounter this effect in later sections. In the present case, not enough information is available to reduce the uncertainty on the FLUKA upper bound value. It should be noted that we cannot generally expect an accuracy at the 10\% level. We trust the electro-thermal MQED estimate, which depends only on the strand enthalpy and the critical temperature; see ``short-duration'' in Sec.~\ref{QuenchLevels}.

\subsection{Dump on Injection Absorber}\label{sec:Q6}
{\bf \noindent  Experimental setup -- }
In order to further study fast-loss events, a quench test was devised that caused an injected bunch at 450\,GeV to be dumped directly on an injection-protection collimator (TCLIB). At the same time, the individually-powered matching-quadrupole magnet (MQM at 4.5~K) Q6.L8 in the shadow of the collimator was powered at varying current levels. The TCLIB jaws were closed to a gap {\bernhard between the jaws} of $\sim$1\,mm, corresponding to the anti-collision limit {\bernhard which prevents the jaws from touching}, and an offset was applied with respect to the beam centre to intercept the full injected beam. 
An oscilloscope was installed on the MQM magnet for enhanced diagnostics to record voltages across the magnet with higher time-resolution than the QPS system could. 

A first test of this kind was performed in July 2011 \cite{ChiaraIPAC12} using a maximum bunch intensity of 3$\times$10$^{10}$ protons 
and a magnet current of up to 2200\,A. The oscilloscope registered a voltage spike at each injection. An offline analysis showed that the spike amplitude varied linearly with bunch intensity. No correlation was found with the magnet current. A normal transition with subsequent fast recovery could, thus, be excluded.
The test was repeated in February 2013. The bunch intensity was 6.5$\times$10$^{10}$ protons and the current was increased in steps of 500~A until a quench occurred at 2500\,A, which {\bernhard corresponds} to operation at 6\,TeV. The 2011 observations on voltage spikes were confirmed. The mechanism causing the spikes is not fully understood at this point.

{\bf \noindent  Particle-shower simulation -- }The energy deposited in the coils was estimated by means of FLUKA simulations, 
reproducing the actual impact conditions on the collimator. 
The simulations included an accurate representation of 
TCLIB, Q6, TCLIM (a mask upstream of Q6), and corresponding aperture transitions. 
The strength of the quadrupole field was adjusted according to the magnet current applied in the test. 
Owing to the jaw length of 1\,m and the active absorber material that is graphite, 
approximately 90\% of the impacting protons experienced an inelastic nuclear interaction inside the jaws, 
while only 10\% of the incident proton energy were absorbed in the TCLIB. 

Comparison of FLUKA results with BLM data proved impracticable. BLMs in the vicinity of TCLIB and MQM saturated. Further downstream, the agreement was not good; the measured BLM signal was 20 times lower than the simulated one. As mentioned above, data from BLMs on a common power distribution line may be unreliable if a large number of BLMs reach saturation. 

\begin{figure}[t]
\centering
\includegraphics[width=0.5\textwidth]{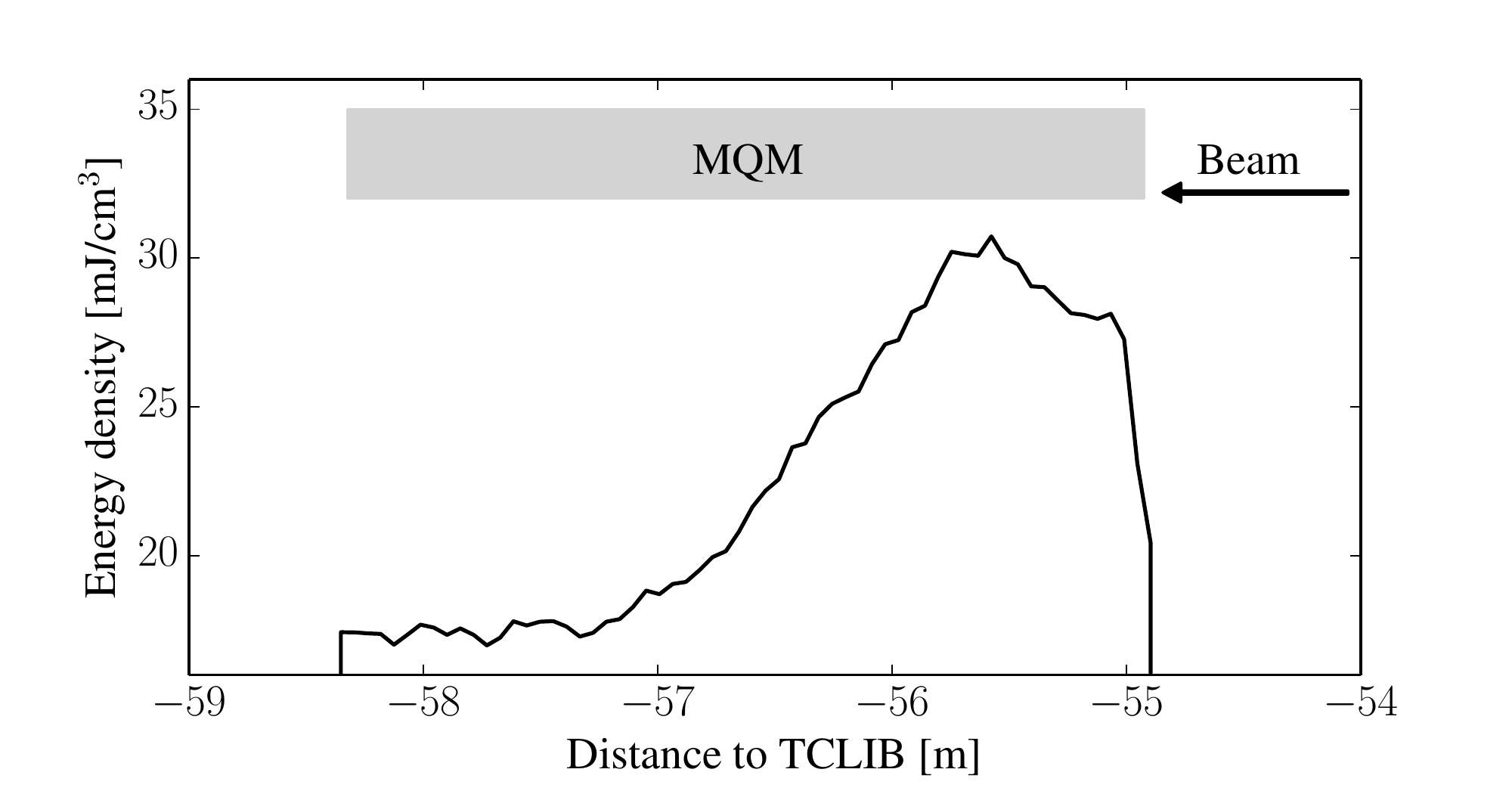}
\caption{FLUKA simulated peak energy density deposited in the coil during the short-duration collimation quench test. The gray box indicates the magnet.}
\label{fig:q6:long}
\end{figure}

\begin{figure}[t]
\centering
\includegraphics[width=0.4\textwidth]{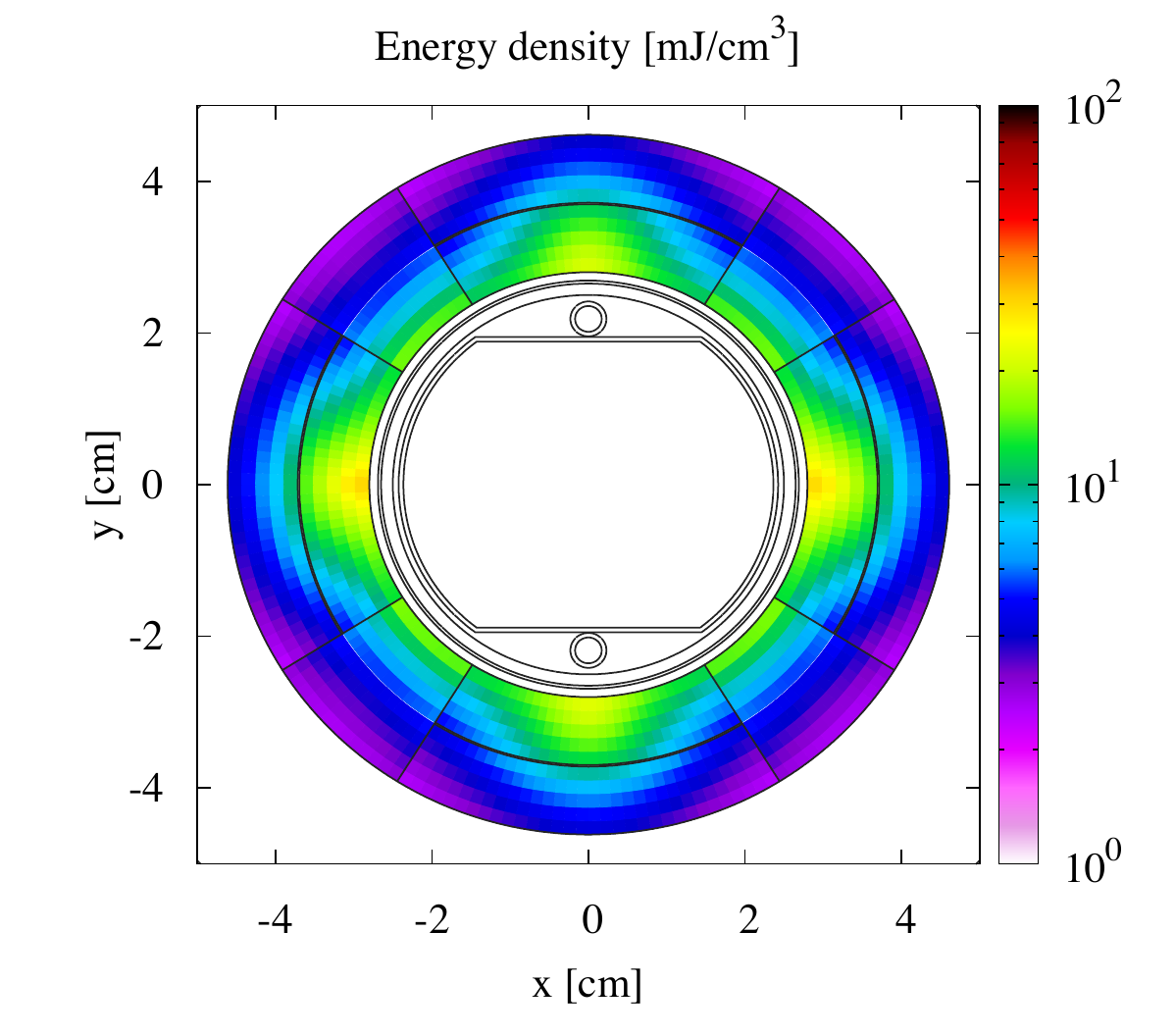}
\caption{Simulated transverse energy density distribution for the short-duration collimation quench test in the Q6.L8 coils (2500\,A) at the position where the maximum energy deposition occurs. 
}
\label{fig:q6:transv}
\end{figure}

For the quench event at 2500\,A 
the FLUKA model predicted a maximum energy density of 31\,mJ/cm$^3$ on the magnet coils. 
A similar peak energy density of 29\,mJ/cm$^3$ was deposited at the lower current of 2000\,A, when no quench was observed; see Tab.~\ref{tab:q6:analyses}. Statistical errors are less than 2\%. Figure~\ref{fig:q6:long} shows that the {\bernhard simulated} losses were distributed over the length of the magnet, with a maximum in the straight part of the magnet. Owing to the larger horizontal elongation of the beam, the maximum energy density was registered in the horizontal plane on the inner coil diameter. This is illustrated in Fig.~\ref{fig:q6:transv}, showing the transverse energy density distribution in the inner and outer layers of the MQM coils. 

{\bf \noindent Electro-thermal simulation -- }
The results of FLUKA and of electro-thermal analyses are shown in Tab.~\ref{tab:q6:analyses}. The losses being instantaneous, helium cooling does not play a role in the electro-thermal model. As above, the MQED is calculated directly from the effective heat capacity of the strand. Comparison between the lower bound for MQED given by FLUKA simulations at 2000~A, and the corresponding electro-thermal MQED estimates shows a mismatch by a factor of roughly 1.5. 

{\bf \noindent Discussion -- }
The quench test was to serve as a bench-mark for the FLUKA model and the electro-thermal model. It is unfortunate that {\bernhard the BLM data was saturated}. In a future experiment, BLM types with higher dynamic range should be used. Nonetheless, from the result we can learn about the necessary safety factors that may need to be applied in the calculation of fast-loss BLM thresholds which are based on similar FLUKA models based on loss scenarios for which BLM data does not yet exist for validation.
\begin{table}[!t]
\caption{Quench-level comparison; FLUKA bounds and the electro-thermal MQED estimate for the short-duration collimation quench test.}
\label{tab:q6:analyses}
\centering
\begin{tabular}{ccc}
\hline
 Current&{ Particle Shower }& Electro-Thermal \\
&Calculation & Estimate\\
\,[A]& [mJ/cm$^3$]& [mJ/cm$^3$]\\
\hline
  2000  &  $>$29  &20 \\
2500&$\leq$31&16\\
\hline
\end{tabular}
\end{table}

\section{Intermediate-Duration Losses}\label{milli}

LHC operation in the years 2010-2013 was affected by a phenomenon of millisecond-duration beam losses. The time-structure of the losses, as observed by the BLM system, was approximately Gaussian. These losses are suspected to be provoked by dust particles getting in the way of the beams \cite{UFO1,UFO1b, UFO2,tobias_thesis}. 
Dust particles can be encountered anywhere around the ring. Statistically, the most frequent {\bernhard single location is at} the injection kickers, affecting mainly the wide-aperture quadrupole (MQY at 4.5~K) Q5. {\bernhard Falling dust particles somewhere in the arcs' main bending (MB) and main quadrupole (MQ) magnets, covering tens of kilometers of the LHC, are expected to be the most critical type of beam losses.} During LS1, one out of three BLMs on the short straight sections in the arc {\bernhard has been} relocated to improve the detection of losses related to dust particles {\bernhard in the arcs}. No quench was provoked during Run 1 and no correlation has been found between beam energy and the occurrence of dust particles. However, the MQED after LS1 is {\bernhard estimated} to be 2-4 times smaller, while the energy deposition due to beam interaction with dust particles is expected to be 2-3 times higher \cite{ref-Anton}. This makes dust particles a prime candidate for beam-induced quenches after LS1. New BLM thresholds have to be determined for the new BLM locations based on the numerical models and test results presented in this section. 

{\bernhard Other beam-loss scenarios of intermediate-duration include sudden current variations in magnet circuits, or losses on collimators at certain stages of beam operation (end of ramp, squeeze). Fast changes in magnet currents, in particular on certain warm magnet circuits \cite{Andres_PhDthesis}, are intercepted by a system of fast magnet-current monitors (FMCMs). The BLM system can intervene as a second line of protection, dumping the beam after the machine-protection system's design response time of 270\,$\mu$s or 3 turns. }

Two experiments were designed to investigate the quench level for intermediate-duration losses: the wire-scanner quench test \cite{WSqTest} and the orbit-bump quench test. 

\subsection{Wire-Scanner Quench Test}\label{sec:wire}
{\bf \noindent Experimental setup -- }One way to generate millisecond losses of roughly Gaussian shape in a controlled way is to use the wire scanner as a source of beam loss. Such an experiment was performed in November 2010 using the wire scanner installed on {\bernhard Beam 2}. This beam was chosen because the collimation region downstream of the wire scanner prevents potential propagation of the losses around the ring.
\begin{figure}[tb]
   \centering
   \includegraphics*[width=0.5\textwidth]{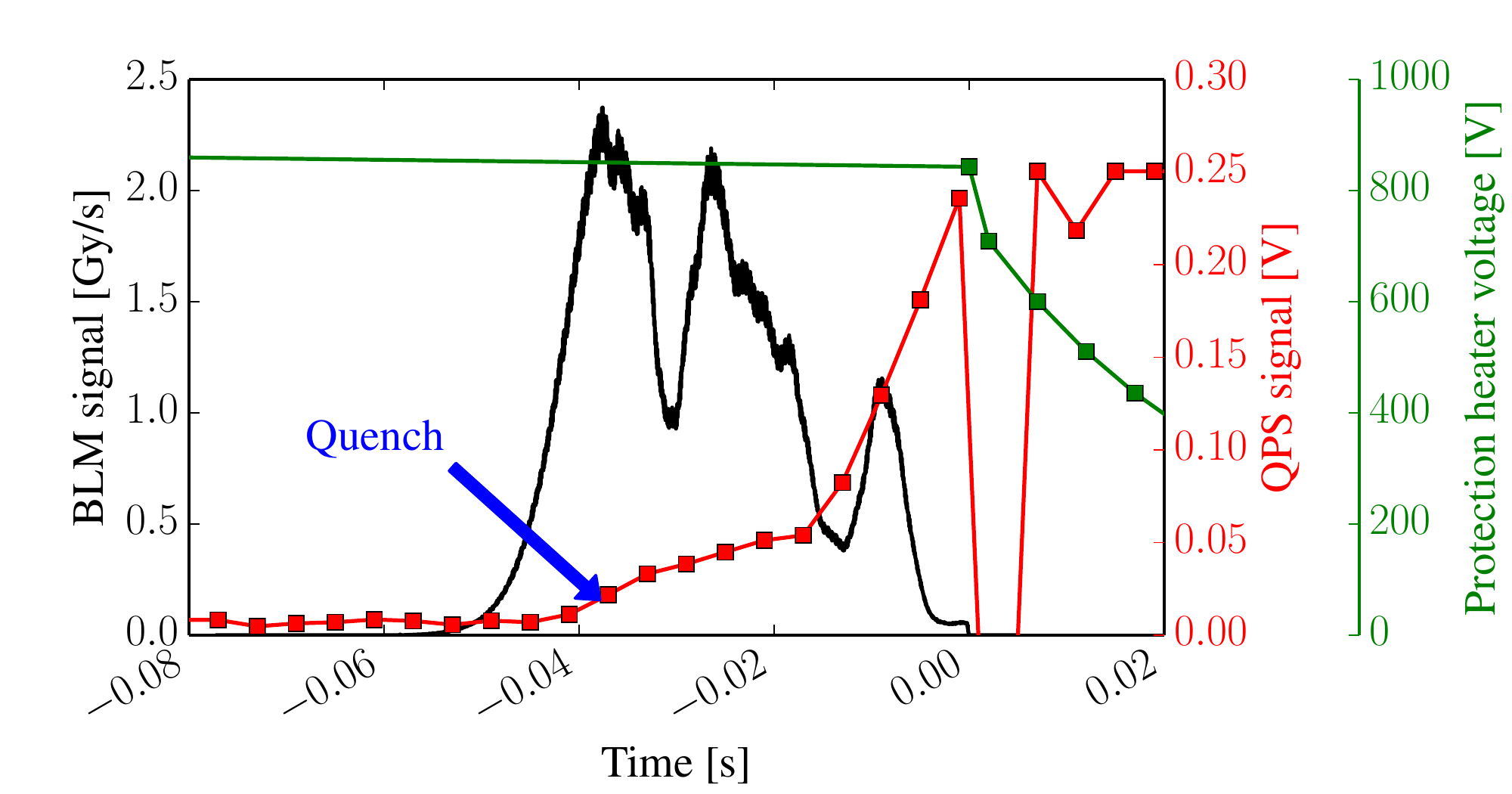}
   \caption{BLM (black) and QPS (red) signals registered during the orbit-bump quench test with intermediate loss duration. A drop in QPS heater-voltage indicates heater firing (green), which was synchronized with the moment of beam dump. The QPS signal measures resistive voltage in a magnet. The 0.1-V intercept of the QPS signal was timed to precede the heater firing by 10 ms, i.e., the QPS evaluation time. }
   \label{WirefastLossesPM}
\end{figure}

The beam intensity was $N_{\rm p}=\rm 1.53\times 10^{13}$ protons contained in 144 bunches at a beam energy of 3.5~TeV. The wires in the scanners are made of carbon fibre with a diameter of $d_{\rm w}=30~\mu{\rm m}$. They perform a linear movement with a nominal speed of \hbox{1 m/s}. During the experiment the speed of the wire $v_{\rm w}$ was gradually decreased with each scan with the sequence: $\rm 1,~0.75,~0.5,~0.37,~0.3,~0.25,~0.2,~0.15,~and~0.05~m/s$, when finally a quench occurred. 

The magnet which quenched was a separation dipole D4.L4 (MBRB type at 4.5~K and 3070~A) placed about 33 meters downstream of the wire scanner. In the same cryostat, there is a quadrupole magnet Q5.L4B2 (MQY type at 4.5~K and 1094~A), which had also been a potential candidate for quenching. Eight BLMs were installed on these magnets.

The quench of the D4 triggered an acquisition of signal buffers, which are presented in Fig.~\ref{WirefastLossesPM}. The non-gaussian shape of the loss registered by the BLMs suggests that the wire movement was not linear and that vibrations occurred together with wire sublimation. Similar behavior of the wires in extremely intense beams was observed before \cite{HB2010_scanners}. Indeed the investigation of the wire after the experiment revealed sublimation of 50\% of wire diameter in the location of the beam impact. The slow rise of the QPS signal is indicative of a quench due to an energy deposition close to the quench level in the magnet ends in the low-field region of the coil.

A precise determination of the moment of quench is difficult due to the long sampling intervals of the QPS data and the uncertainty in the synchronisation of individual signals. These uncertainties affect the particle-shower simulation of the energy deposited in the coil at the moment of quench, as well as the electro-thermal MQED estimate. 

{\bf \noindent {\bernhard Particle-shower simulation} -- }For a normal scan of a gaussian beam the amount of protons passing through the wire $N_{\rm w}$ can be expressed by \cite{AntonFLUKA}
\begin{equation}
N_{\rm w} = N_{\rm p} \cdot f_{\rm LHC} \cdot d_{\rm w} / v_{\rm w}
\label{eq:nwire}
\end{equation}
where $\rm f_{\rm LHC} = 11~kHz$ is the revolution frequency of protons in the LHC, $v_{\rm w}$ the wire velocity, $d_{\rm w}$ the wire diameter, and $N_{\rm p}$ the number of protons in the beam. In order to estimate the actual number of protons interacting with the wire in the last, irregular scan, we use an unexpected increase of the integrated BLM signal. For all preceding scans we had found that the product of wire-speed and BLM signal $S_{\!\rm BLM}$ was constant, $S_{\!\rm BLM} \cdot v_{\rm w} = {\rm const}.$ Moreover, with $N_{\rm w}$ from  Eq.~(\ref{eq:nwire}), we had found that $N_{\rm w}$ was proportional to the BLM signal, $N_{\rm w} \propto S_{\!\rm BLM}$. For the last scan, $S_{\!\rm BLM}\cdot v_{\rm w}$ was 30\% higher than expected. This finding translates to an estimated $N_{\rm w}$ of $1.3\times 10^{14}$ protons. 

\begin{figure}[tb]
   \centering
   \includegraphics*[width=0.5\textwidth]{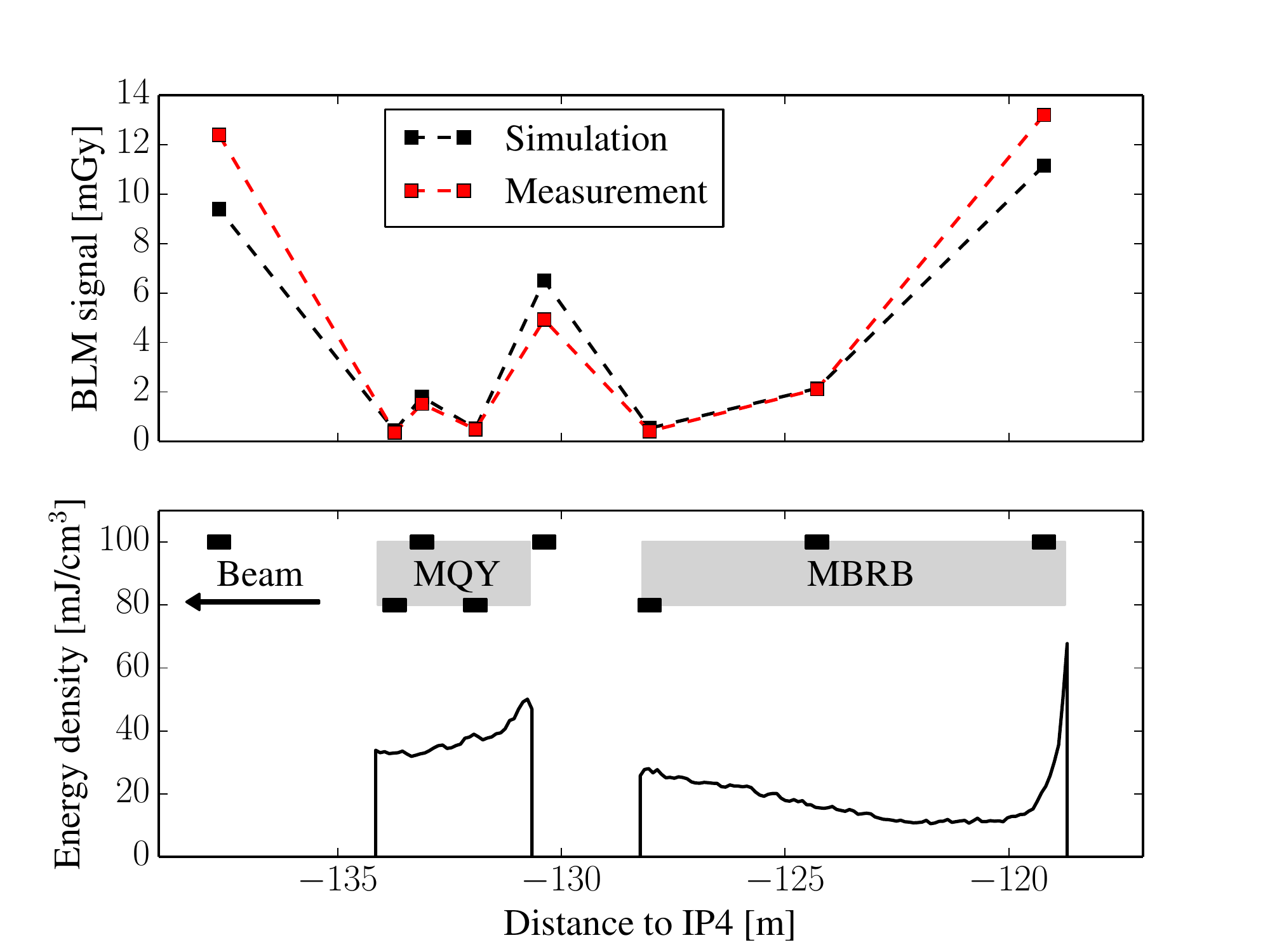}
   \caption{Up: Comparison of integrated BLM signal accumulated during the wire-scanner quench test and simulated with FLUKA. Down: FLUKA simulated peak energy density deposited in the coils over the entire event. Gray boxes indicate magnet cold masses, black boxes indicate the locations of BLMs. }
   \label{F:3_wsq_blm}
\end{figure}
\begin{figure}[t]
\centering
\includegraphics[width=0.4\textwidth]{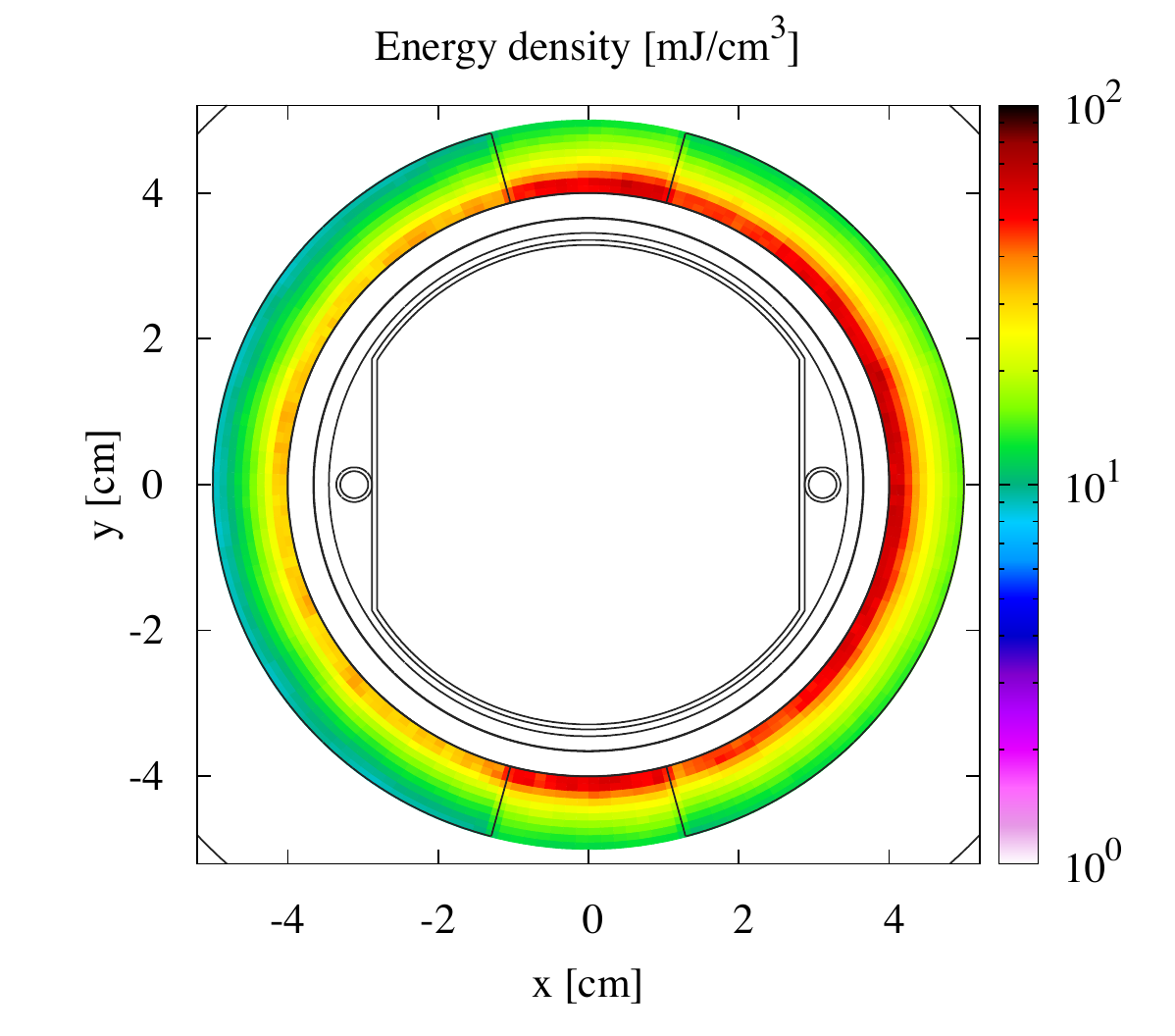}
\caption{Simulated transverse energy density distribution {\bernhard from FLUKA}, integrated over the event, in D4.L4 coils at the
location where the maximum energy deposition occurs during the intermediate-duration orbit-bump quench test.}
\label{fig:ws:transv}
\end{figure}
The agreement in shape and amplitude of FLUKA simulations with BLM data (see Fig.~\ref{F:3_wsq_blm} up) was very good, thus, vindicating the above considerations. The uncertainties affect the calculation of the energy deposited in the coil at the moment of quench. In the MBRB, the peak energy deposition occurred in the coil end; see {\bernhard Fig.~\ref{F:3_wsq_blm} (down)}. This makes a precise determination of the locally deposited energy in the magnet coil difficult. 
FLUKA results are shown in Tabs.~\ref{tab:wire:analyses1} and~\ref{tab:wire:analyses2}. The corresponding transverse energy-deposition in the MBRB coils is shown in Fig.~\ref{fig:ws:transv}.



{\bf \noindent Electro-thermal simulation -- }MQED estimates for the beam-loss scenario presented by the wire-scanner quench test are affected by the uncertainty on the moment of quench in the same way as FLUKA simulations, thus, adding an uncertainty to $N_{\rm q}$, the number of protons lost until the MBRB magnet quenched for $v_{\rm w}=0.05$~m/s. A synchronisation of signals as shown in Fig.~\ref{WirefastLossesPM} means that $N_{\rm q}/N_{\rm w}=30$\%; if the quench occurred 5~ms later (recall that 5~ms is the QPS sampling rate) we find $N_{\rm q}/N_{\rm w}=45$\%. For the quench test with $v_{\rm w}=0.15$~m/s no quench occurred. The BLM signals are not available, so we assume a horizontal beam distribution with $\sigma_{\rm h}=300~\mu$m and, consequently, a Gaussian time distribution with $\sigma_{\rm t}=\sigma_{\rm h}/v_{\rm w}=2~$ms. Since the FLUKA model of the MBRB does not feature an accurate geometrical model of the coil ends, the location of quench and, hence, the magnetic field in the location of quench are not well known. Moreover, due to the filling of gaps in the magnet ends with putty, the cooling conditions in the ends are little known. Tables~\ref{tab:wire:analyses1} and~\ref{tab:wire:analyses2} show the simulation results. The lower MQED estimate corresponds to a situation with no helium cooling and higher local magnetic field, whereas the higher MQED value considers the standard model as it would be applied in the straight section of the magnet. The maximum in the loss distribution inside the MQY coils is 10~cm inside the magnet coil. We therefore assume standard cooling and field conditions in the MQY. The fact that the magnets are operated at 4.5~K means that the helium inside the cable plays a lesser role than, for instance, in the orbit-bump quench test in Sec.~\ref{sec:fastADT}. 
\begin{table}[!t]
\caption{Quench-level comparison; FLUKA estimate and electro-thermal MQED estimate in the MBRB coil for the wire-scanner quench test. {\bernhard Lower values in electro-thermal estimates correspond to a reduced cooling model; upper values correspond to the standard cooling model.} }
\label{tab:wire:analyses1}
\centering
\begin{tabular}{cccc}
\hline
 $v_{\rm w}$ &$N_{\rm q}/N_{\rm w}$&{ P. Show.}& El.-Therm. \\
 \, [m/s]& [\%]&[mJ/cm$^3$]& [mJ/cm$^3$]\\\hline
 0.15&n/a&$>$18&26-37\\
0.05&30&20&25-35\\
0.05&45&30&26-42\\
\hline
\end{tabular}
\end{table}
\begin{table}[!t]
\caption{Quench-level comparison; FLUKA bound and the electro-thermal MQED estimate in the MQY coil for the wire-scanner quench test. }
\label{tab:wire:analyses2}
\centering
\begin{tabular}{ccc}
\hline
 $v_{\rm w}$ &{ Particle Shower}& Electro-Thermal \\
 &Calculation&Estimate\\
 \, [m/s]& [mJ/cm$^3$]& [mJ/cm$^3$]\\
\hline
0.05&$>$50&52\\
\hline
\end{tabular}
\end{table}

{\bf \noindent Discussion -- } 
The wire scanner is an ideal device to generate milli-second-duration losses with a Gaussian loss profile. Its position in the ring allows to create losses in the D4 magnet that are sufficiently intense to make the magnet quench. Unfortunately, the relative position of wire scanner and D4 is such that the peak losses occur in the magnet ends, making an accurate simulation rather difficult. The test is not likely to be repeated due to the limited supply of spare magnets for the D4. A similar experiment on a magnet cooled to 1.9~K would be desirable. The losses, in that case, would have to be considerably higher (see the following section), and the wire scanner would have to be fitted for the purpose. In any case, any experiment in the intermediate-duration regime should record BLM and QPS signals with an oscilloscope to avoid timing issues in the analysis. 

\subsection{Orbit-Bump Quench Test }\label{sec:fastADT}

{\bf \noindent Experimental setup -- }The LHC transverse damper (ADT)~\cite{adt} can be used not only to {\bernhard damp} beam oscillations but also to excite the beam {\bernhard by so-called operation in sign-flip mode, which gives kicks to selected bunches}.
The preparation of the beam excitation procedure for the orbit-bump quench test is described in \cite{thisProc}. First, a three-corrector orbit bump was applied in the horizontal plane around the main quadrupole MQ.12L6 (at 1.9~K). Second, the tune kicker (MKQ) kicked the bunch horizontally. Third, with a delay time of 1~ms or 11~turns, the horizontal ADT started the excitation of a single bunch in sign-flip mode. Two attempts were made. Both times the entire bunch  {\bernhard containing} $N_\mathrm{p}={\bernhard 4}$$\times$10$^8$ protons in the first try and $N_\mathrm{p}=8.2$$\times$10$^8$  in the second try was lost into the magnet. The second attempt resulted in a quench in the magnet. A particular challenge lay in the measurement of the beam intensity and the emittance of bunches with several $10^8$ protons per bunch, which is more than ten times below the design sensitivity of the LHC beam instrumentation.

\begin{figure}[t]
   \centering
   \includegraphics*[width=0.5\textwidth]{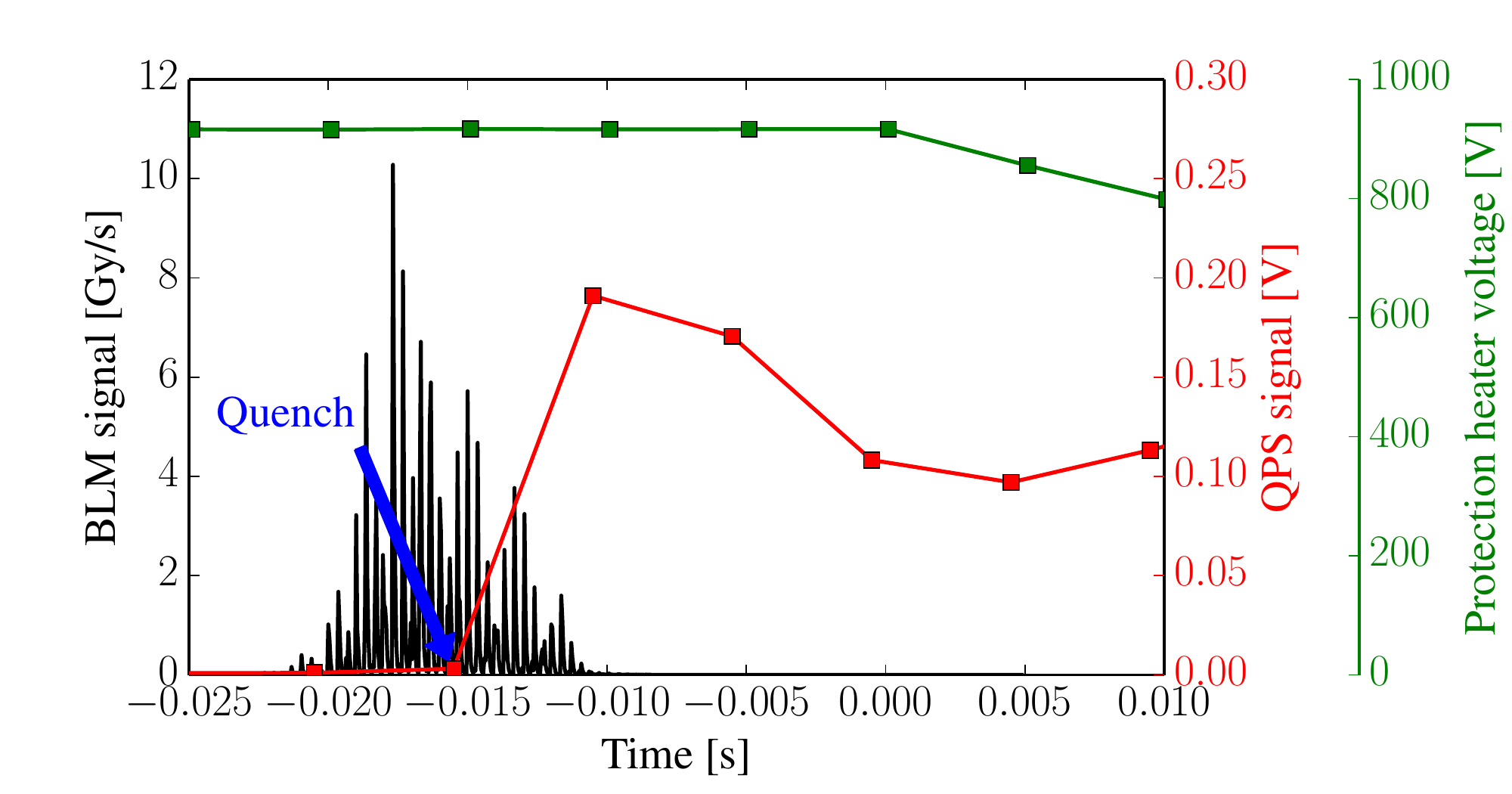}
   \caption{BLM (black) and QPS (red) signals registered during the orbit-bump quench test with intermediate loss duration. A drop in QPS heater-voltage indicates heater firing (green), which was synchronized with the moment of beam dump. The QPS signal measures resistive voltage in a magnet. The 0.1-V intercept of the QPS signal was timed to precede the heater firing by 10 ms, i.e., the QPS evaluation time.
    }
   \label{ADTfastLossesPM}
\end{figure}

In Fig.~\ref{ADTfastLossesPM} the recorded BLM and QPS signals of the second attempt are presented. The total duration of losses was about \hbox{10 ms}, with loss spikes roughly every four revolutions of the excited {\bernhard bunch}. The determination of the precise moment of quench suffers from the low sampling rate (5~ms) of QPS data and the imperfect synchronisation of BLM, heater, and QPS signals. For this purpose, an oscilloscope was installed, which, however, stopped operating during the first attempt, possibly due to radiation issues. For MQED estimates, we assume that the quench occurred after about \hbox{5 ms} when about {\bernhard$N_\mathrm{q}=5.3$$\times$10$^8$} protons were lost; see caption of Fig.~\ref{ADTfastLossesPM}. 

The particular shape of the QPS signal allows for two different interpretations. The QPS signal is the difference between two voltages, each measured across two poles of the affected quadrupole. Since the QPS signal has a local minimum at $\sim$+5~ms (see Fig.~\ref{ADTfastLossesPM}), the magnet may be recovering to a superconducting state before the protection heaters become effective. Quench-recovery in an MQ at \hbox{{\bernhard 4 TeV}} and the corresponding magnet current is, however, very unlikely. Alternatively, the second voltage signal may be ``catching up'' to the first signal, due to an almost symmetric quench development across the two pairs of voltage taps. For such cases, an additional layer of protection is provided by the symmetric quench protection system that compares voltages across apertures of adjacent magnets. 

\begin{figure}[t]
   \centering
   \includegraphics*[width=0.5\textwidth]{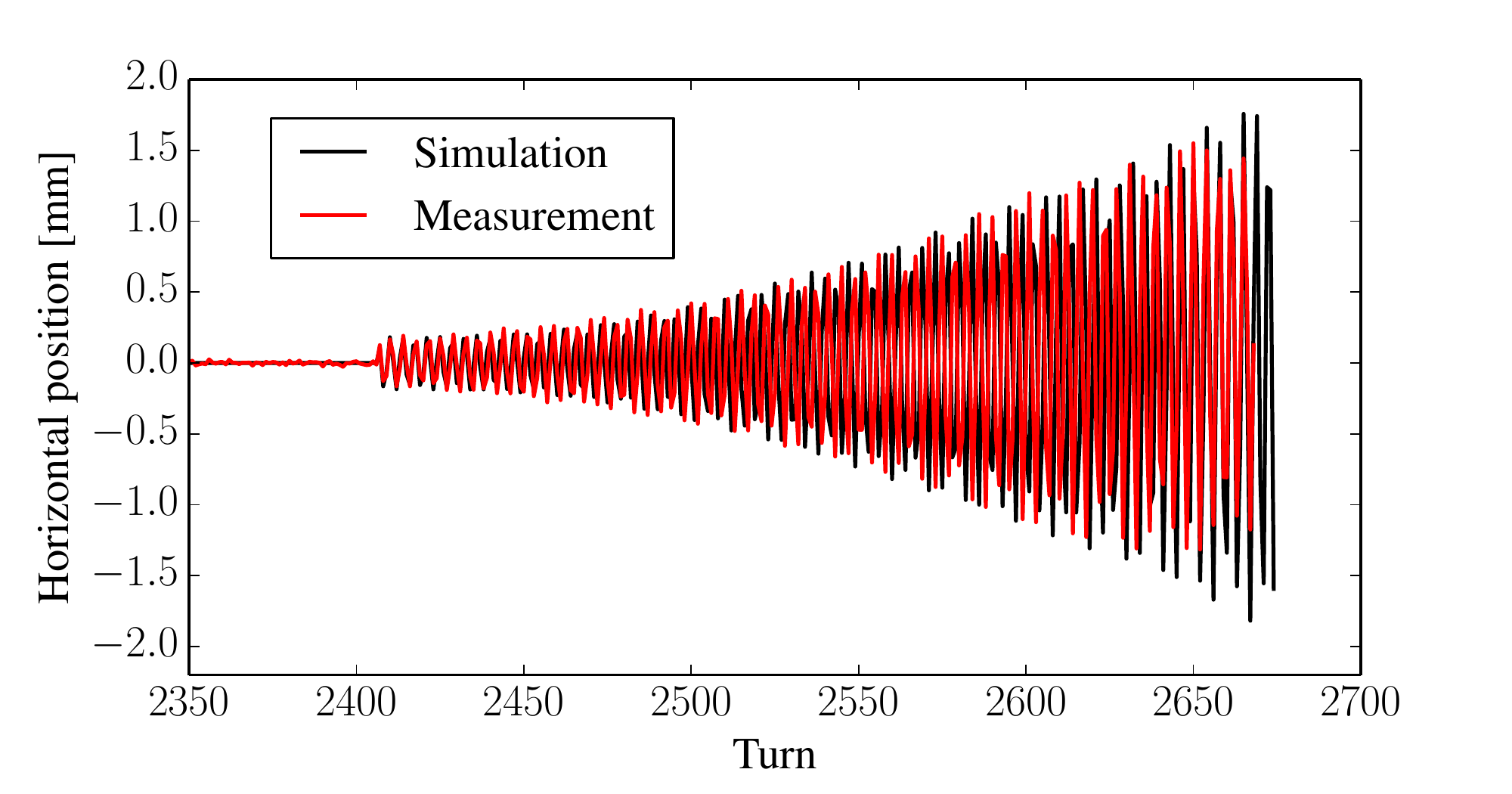}
   \caption{Comparison of MAD-X simulation and data collected by a beam-position monitor during the intermediate-duration orbit-bump quench test.}
   \label{F:4_madx}
\end{figure}

\begin{figure}[!t]
\centering
\includegraphics[width=0.5\textwidth]{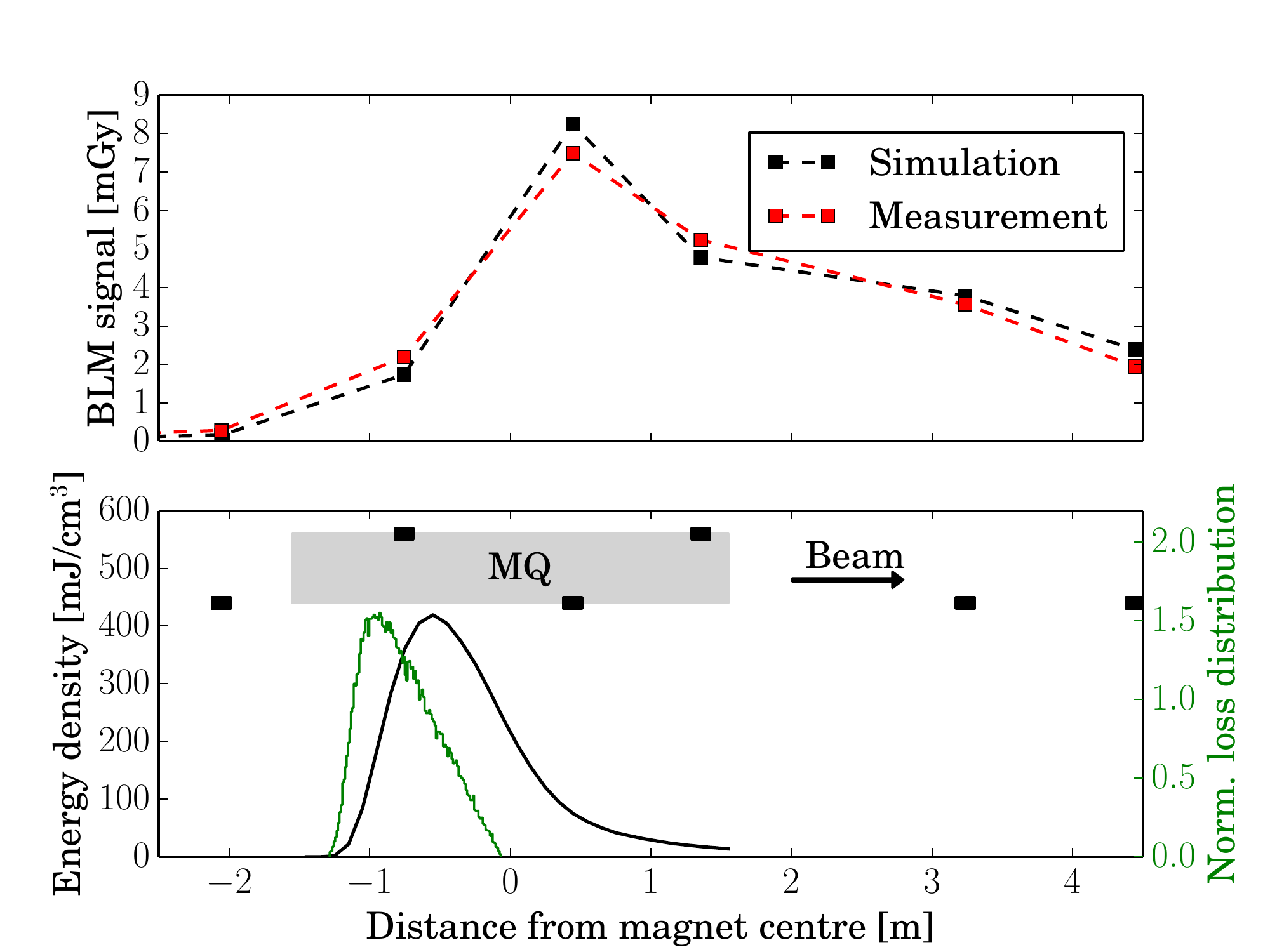}
   \caption{Up: Detail of the comparison between the BLM signal, accumulated during the intermediate-duration orbit bump quench test, and the simulated signal from FLUKA. Down: FLUKA simulated peak energy density deposited in the coils (black) and MAD-X histogram of protons lost on the beam-screen (green), normalised to the total number of simulated lost particles. The gray box indicates the magnet, black boxes indicate the locations of BLMs.}
   \label{fig:adtfast:blms}
\end{figure}

{\bf \noindent Particle tracking -- }This quench test required a much more detailed particle-tracking study than the single-turn and fixed-target losses described so far. Extensive tracking studies with MAD-X \cite{MADX} have been performed to model
the spatial as well as the time distribution of losses on the beam screen from the excited bunch, \cite{vera_ipac14}. 
In order to fully describe the experimental conditions, the simulations strictly followed the chronology of the experiment: the orbit bump was followed by an MKQ kick, and the ADT excitation. The BPM data in the position of the ADT was used for tuning the strengths and directions of the MKQ and ADT-kicks in the MAD-X model. In the simulation, the ADT kick is treated as a sine function with growing amplitude for the first 100 turns, and constant amplitude thereafter, corresponding to a saturation of the ADT kick strength.
In Fig.~\ref{F:4_madx} the simulated beam position at the BPM 
is compared to the experimental data. 
The time, position, impact angle, and energy of every particle touching the aperture is stored. The results of the tracking simulations are used as input for FLUKA particle shower simulations; {\bernhard see} Fig.~\ref{fig:adtfast:blms} (down).

Parametric studies revealed that the largest uncertainties in the particle-tracking results are due to the imperfect knowledge of  tune and  beam profile, moments before the beam losses occurred. The maximum in the spatial loss distribution may be 20\% lower if the tune after establishment of the three-corrector bump was 64.268 rather than the nominal 64.274 or, alternatively, if the beam was twice as wide. The amplitude of the orbit bump, when increased, shortens the overall longitudinal loss distribution; however, since the maximum of the loss distribution remains unchanged, this variation leaves BLM signals and the peak energy-distribution in the coils unchanged.
 
Particle-tracking results for orbit-bump scenarios were systematically checked for their sensitivity to {\bernhard tiny} discontinuities in the beam-screen surface, e.g., a region of elevated surface roughness. This type of imperfection was modeled as a 20-cm-long aperture restriction of 30~$\mu$m height. The effect of the restriction depends largely on how fast the beam is driven towards the aperture. To generate losses in the millisecond time-range, the combined {\bernhard MKQ} and ADT kicks cause a relatively wide loss distribution, with a maximum towards the beginning of the magnet; compare Fig.~\ref{fig:adtfast:blms} with Fig.~\ref{fig:ssqt:blms}, {\bernhard where the latter shows a similar event but with slow beam blow-up over 20 seconds}. The impact angle varies linearly along the length of the quadrupole \cite[Fig.~1]{vera_ipac14}. As a consequence, an aperture restriction at the beginning of the magnet shields a short downstream section from impacting protons, whereas a restriction towards the magnet centre can effectively curtail a loss distribution. From this reasoning, and the loss distribution shown in  Fig.~\ref{fig:adtfast:blms}, it follows that orbit-bump tests with strong kicks are not highly sensitive to small aperture restrictions. Moving the restriction along the magnet causes either a shorting at the end, or a prolongation at the beginning of the distribution; however the change in overall results (including the subsequent particle-shower simulation) does not exceed 10\% for the assumed roughness height of 30~$\mu$m.


While the spatial loss distribution showed low sensitivity to parameter changes, the time structure of losses varied more strongly. Since the beam was excited for a short time only, its blow-up could be neglected, besides excitation happened too fast in order to allow for cutting the phase-space ellipse along its perimeter. Therefore the envelope of the loss-peaks had a Gaussian-like shape, reflecting the shape of the beam profile. The frequency of the loss peaks strongly correlates with the tune. The duration of the loss depends on the beam size and on the orbit bump amplitude. An accurate reproduction of the time structure observed in the BLM signals could not be achieved. The likely reasons are  the above mentioned uncertainties in the tune and the beam profile measurements,
as well as higher-order effects.

{\bf \noindent Particle-shower simulation -- }The MAD-X simulations predict a spatial loss distribution which is restricted to a 1.2-m-long area upstream of the longitudinal magnet center. The impact angles of protons on the magnet beam screen gradually decrease from the beginning towards the {\bernhard centre of the magnet}, owing to the focussing quadrupole field. The MAD-X loss distribution was integrated over time, and used as an input for FLUKA shower calculations \cite{nikhil_ipac14}. Figure~\ref{fig:adtfast:blms} (up) compares simulated and measured signals for a string of BLMs positioned along the cryostat of the impacted quadrupole magnet. For BLMs located downstream of the assumed loss location, the agreement between simulation and measurement is found to be better than 30\%. 

Parametric studies involving both, FLUKA and MAD-X, have been carried out. The peak energy deposition in the coil was shown to correlate with the peak in the geometrical distribution of lost particles on the beam screen. The energy deposition in the location of the BLMs was considerably less sensitive. Sharp peaks and more drawn-out distributions on the beam screen produce in the BLMs nearly identical signatures, owing to the very forward direction of the particle shower. 

\begin{figure}[t]
\centering
\includegraphics[width=0.4\textwidth]{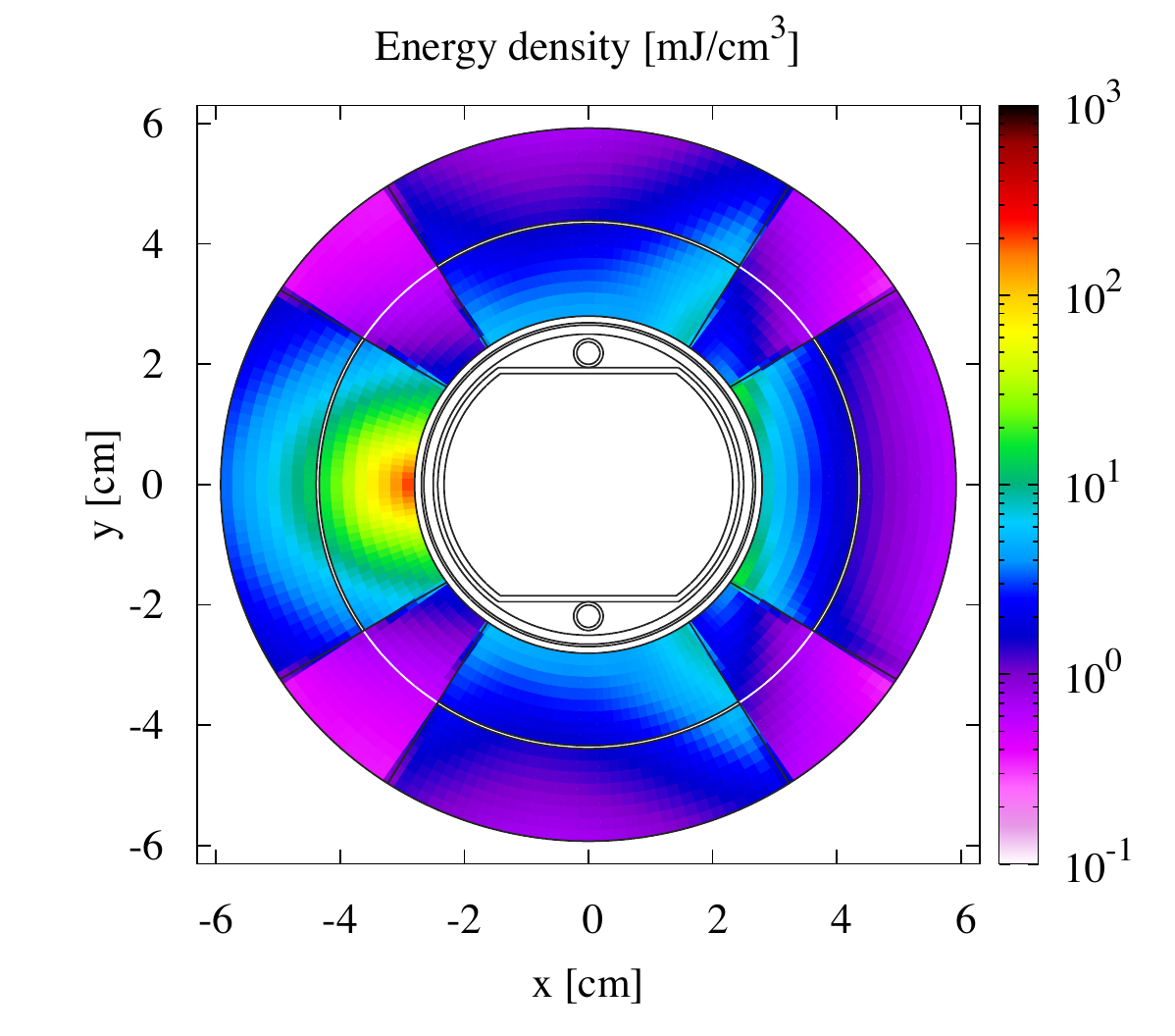}
\caption{Simulated transverse energy density distribution {\bernhard from FLUKA} in MQ.12L6 coils at the
location where the maximum energy deposition occurs during the intermediate-duration orbit-bump quench test. Results correspond to {\bernhard 4$\times$10$^{8}$} protons impacting on the magnet beam screen. 
}
\label{fig:adtfast:transv}
\end{figure}

Figure~\ref{fig:adtfast:transv} illustrates the transverse energy density profile at the position of the maximum energy deposition. The displayed distribution corresponds to a cumulative loss of {\bernhard $N_{\rm p}=4$$\times$10$^8$} protons. Lower bounds are imposed by the intensity of bunches used in tests without quench,  {\bernhard$N_{\rm p}=4$$\times$10$^8$}. For a bunch intensity of $N_{\rm p}=8.2$$\times$10$^8$ protons, the experiment resulted in a {\bernhard quench. Owing} to the limited time resolution of the QPS signals, the number $N_{\rm q}$ of protons impacting on the beam screen up to the moment of quench cannot be determined conclusively. For {\bernhard $N_{\rm q}=5.3$$\times$10$^8$} protons lost, corresponding to the assumption that the magnet quenched after 5~ms of losses (see Fig.~\ref{ADTfastLossesPM}), the estimated maximum energy density is about {\bernhard 265\,mJ/cm$^3$}; see also Tab.~\ref{tab:fastADT:analyses}. An upper bound to the quench level is given by $N_{\rm q}= N_{\rm p}$, which results in an MQED of 405\,mJ/cm$^3$.

{\bf \noindent Electro-thermal simulation -- }
The radial distribution of losses determined by FLUKA is normalised to the maximum value and used for the loss profile in the electro-thermal simulation. For the time distribution, the BLM signals were normalised to their maximum value and truncated at the presumed moment of quench. Simulation results are shown in Tab.~\ref{tab:fastADT:analyses}. We find that the electro-thermal model appears to underestimate the quench level in this regime substantially. Several effects have been discussed in this context: A thin helium film all around the affected strand could increase the cooling capability in the first instances of losses considerably; current re-distribution could delay the measurable resistive signal, even though the longitudinal peak is relatively broad for this effect to play a decisive role (see Fig.~\ref{fig:adtfast:blms}); nucleate boiling has been shown, albeit in a semi-infinite bath, to be highly effective on very short time ranges \cite{nb}. Numerical studies show that if we allow the cooling model to fall back to the nucleate-boiling regime after each short loss peak, the observed values could be roughly reproduced. In the absence of a comprehensive fluid-dynamic model, however, this observation is merely the ground for speculations. This enhanced nucleate-boiling regime was not used for the values presented in Tab.~\ref{tab:fastADT:analyses}. We need more experimental and theoretical work to arrive at a predictive model of helium-cooling with superfluid helium in Rutherford-type cable in the intermediate-duration loss regime. Note that the wire-scanner quench test, which was carried out on a magnet operated at 4.5~K, did not show this kind of underestimation. 

\begin{table}[!t]
\caption{Quench-level comparison; FLUKA calculations and electro-thermal estimates of the MQED in the MQ coil for the intermediate-duration orbit-bump quench test. {\bernhard Lower values in the electro-thermal estimate correspond to a more pessimistic cooling model; both cooling models neglect a potential increase of cooling power by an enhanced nucleate-boiling regime.}}
\label{tab:fastADT:analyses}
\centering
\begin{tabular}{cccc}
\hline
$N_{\rm p}$ &$N_{\rm q}$ &{ P. Show.}& El.-Therm. \\
& & [mJ/cm$^3$]& [mJ/cm$^3$]\\
\hline
  4$\times$10$^8$&n/a& $>$198&61-71\\
8.2$\times$10$^8$&5.3$\times$10$^8$&{\bernhard 265}&50-58\\
8.2$\times$10$^8$&8.2$\times$10$^8$&$\leq$405&70-80\\
\hline
\end{tabular}
\end{table}

{\bf \noindent Discussion -- }
The experiment succeeded in generating losses over several milliseconds that resulted in a magnet quench. The quench occurred in the straight section of the magnet, thus avoiding a problem observed in the wire-scanner quench test. The particular time structure of losses, with peaks roughly every 300~$\mu$s, represents an important deviation from the Gaussian-shape losses due to dust particles. The experiments may, therefore, not be all-together suitable to draw conclusions on limitations due to falling dust particles for LHC operation. As in the case of the wire-scanner quench test, we note that oscilloscope recordings of BLM and QPS signals are mandatory for any future test.

\section{Steady-State Loss Quench Tests}\label{steady}

Steady-state losses are generated by luminosity debris hitting magnets close to the experiments, and by collimators in the cleaning insertions. These losses cannot be avoided therefore they set limits to the machine performance. Residual particle showers from the collimation system constitute a well-defined scenario, directly amenable to experimental testing. The 2011 and 2013 collimation quench tests were performed with protons and ions \cite{Daniel_collProtons,2011:colliqt} without quenching, {\bernhard thus, providing a lower bound on the quench level, i.e., on MQPD}. Five tests using the orbit-bump technique with protons resulted in quenches. Here we present the 2013 collimation quench test with protons in Sec.~\ref{sec:collimation}, and orbit-bump quench tests in Secs.~\ref{sec:dynamic} and ~\ref{sec:slowADT}, which test the quench level in main quadrupoles for increasing and {\bernhard near-constant} power deposition, respectively.


\subsection{Collimation Quench Tests}\label{sec:collimation}

{\bf \noindent Experimental setup -- }During regular LHC operation, the dispersion suppressor magnets (DS) in IR7 are the superconducting elements that are most exposed to beam losses leaking out of the betatron collimation system~\cite{lhc:design_report, coll:assmann_chamonix05,coll:assmann_ecap06,coll:wollmann_ipac10}. These losses, together with the beam lifetime, limit the maximum beam intensity that can be injected. Dedicated collimation quench tests were devised to explore this limit in the DS regions. 

In 2011, two tests were performed, with protons and ions, respectively, at $3.5\,\mathrm{Z\,TeV}$~\cite{2011:collpqt,2011:colliqt}. The main goal was to achieve a loss rate of $500\,\mathrm{kW}$ which is the maximum loss rate the collimation system was designed to intercept~\cite{lhc:design_report}. Beam losses on the collimators were triggered for Beam 2 over $1\,\mathrm{s}$ by crossing the third-order resonance to blow-up the beam. The fraction of particles leaking from the collimators into the DS region, as well as their impact distribution, stayed the same as in standard operation, while the number of lost particles increased significantly. The method allowed to investigate performance limitations due to the DS magnets' quench level. The leakage, however, was not high enough to provoke a quench.

In this paper, we describe the more recent experiment performed in 2013 with protons at {\bernhard $4\,\mathrm{TeV}$}; see also~\cite{2013:collqt}. In order to increase the losses in the DS of IR7 with respect to the 2011 tests, the collimator settings were changed. We adopted the relaxed collimator settings used during the 2011 run, and opened further the secondary collimators in IR7\ifdefined\INCLUDECOLLFIG; see Tab.~\ref{tab:coll_set_2013}\fi. The global effect of these changes was to increase the number of impacts in the DS of IR7 for the same beam loss. 
The optimisation of the settings was the result of the combination of tracking studies with SixTrack~\cite{coll:sixtrack,K2,SixTrack3,BrucePRSTAB}, and a detailed validation during a low-intensity fill before the quench test. 

\ifdefined\INCLUDECOLLFIG
\begin{table}[t]
\caption{Collimator settings used in 2013 collimator quench test expressed in beam sigma size at 4\,TeV, normalised transverse emmittance of $3.5\,\mathrm{\micro m\,rad}$. {\bernhard META: Maybe this is more relevant for the technical note?} \label{tab:coll_set_2013}}
\centering
    \begin{tabular}{ccccc} \hline \hline
    {\bf TCP7}                                   & {\bf TCSG7} & { \bf TCLA7} & {\bf TCSG6} & {\bf TCDQ } \\ \hline \hline
    6.1 & 10.1 & 18.9 & 10.9 & 11.5 \\ \hline
     \end{tabular}
\end{table}
\fi

\begin{figure}[t]
	\centering
		\includegraphics[width=0.5\textwidth]{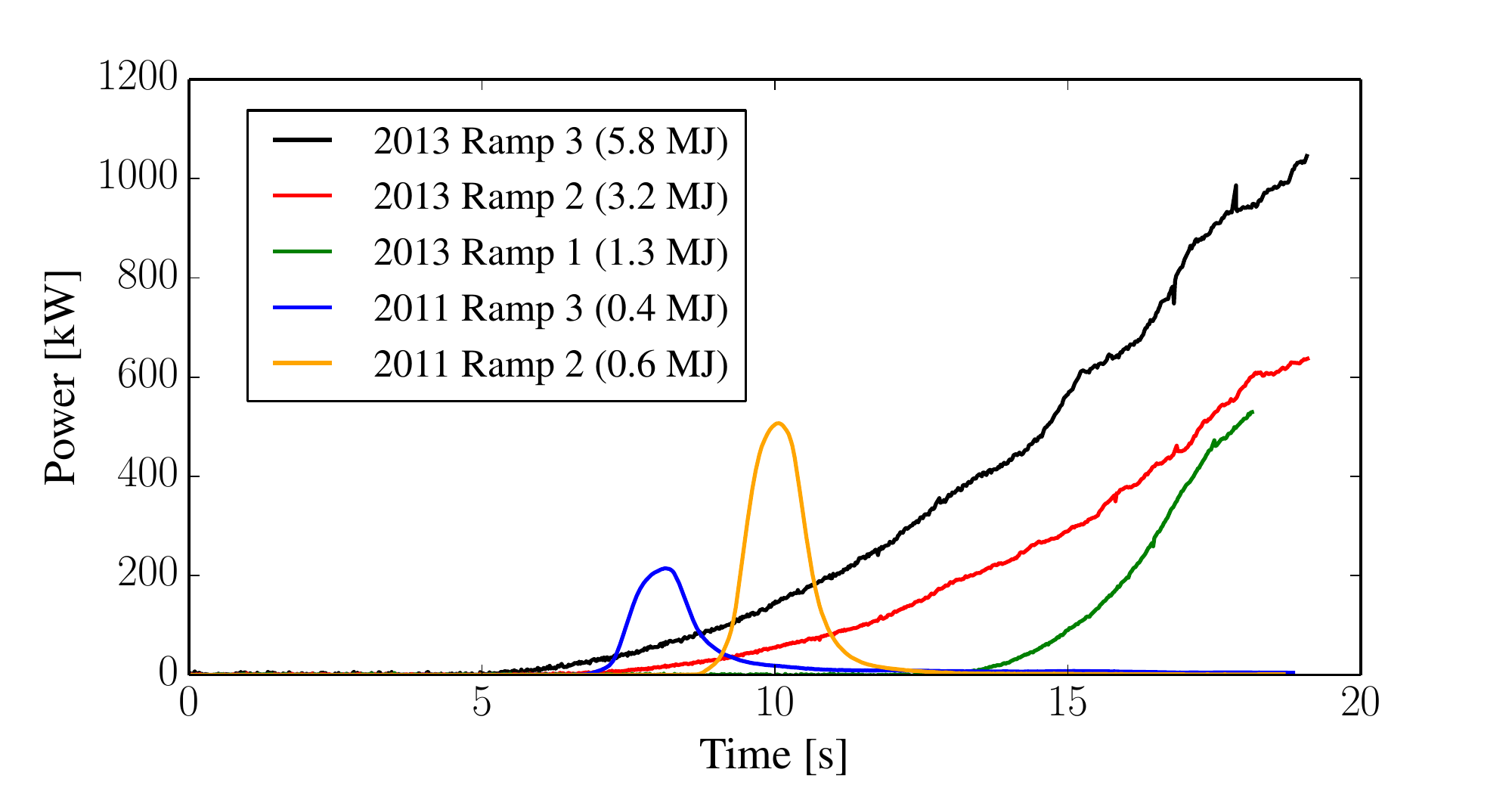}
	\caption{{\bernhard Measured peak power losses} by the beam in the collimation system versus time during the collimation quench tests in 2011 and 2013. In 2013 tests, beams were dumped after achieving the targeted loss rate.}
	\label{fig:coll:power_loss}
\end{figure}

Beam losses on the primary collimator (TCP in IR7) were generated by blowing up the beam with white noise from the ADT~\cite{adt}. This mechanism allowed to generate beam losses that increased continuously over $10\,\mathrm{s}$. Figure\ifdefined\INCLUDECOLLFIG s~\ref{fig:coll:intensity} and\fi~\ref{fig:coll:power_loss} show\ifdefined\INCLUDECOLLFIG\else s \fi the peak power loss during the 2011 and the 2013 tests. Beam losses of up to $1050\,\mathrm{kW}$ were generated in the last ramp with up to $5.8\,\mathrm{MJ}$ impacting on the primary collimator over a few seconds. During that period the BLM signals were monitored. \ifdefined\INCLUDECOLLFIGFigure~\ref{fig:coll:lossmap_ring} shows the BLM signals along the LHC ring at the peak $1\,\mathrm{MW}$ beam loss. The blue lines represent the measured BLM signal at the superconducting magnets. In the left of IR7 ($s=19994\,{\rm m}$), the leakage generated during the test can be seen.\fi As in 2011, no magnet quench occurred. {\bernhard The maximum BLM signals in the cold sector were measured at the position of the main quadrupole MQ.8L7.}

\ifdefined\INCLUDECOLLFIG
	\begin{figure}[t]
		\centering
		\includegraphics[width=0.4\textwidth]{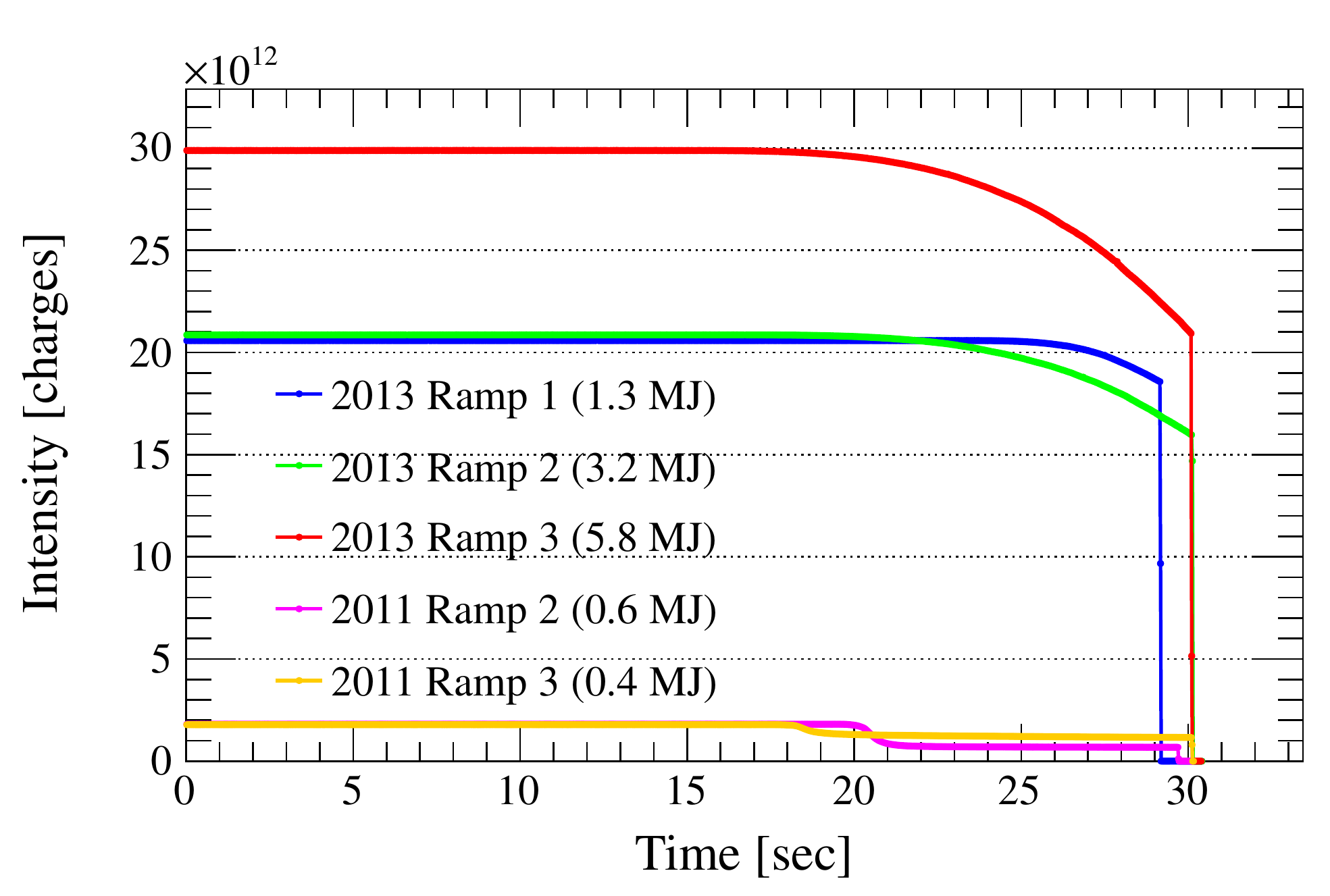}
		\caption{Beam intensity as a function of time during the collimation quench tests in 2011 and 2013.{\bernhard META: 
		Information in this plot seems slightly redundant comparing with Fig. 10.}}
		\label{fig:coll:intensity}
	\end{figure}
\else
\fi

\ifdefined\INCLUDECOLLFIG
	\begin{figure*}[tb]
		\centering
		\includegraphics[width=0.7\textwidth]{20130215_031503_B1_B2_3569_4000GeV_raw_F}
		\caption{Raw signal from BLMs along the LHC ring during $1\,\mathrm{MW}$ beam losses.}
		\label{fig:coll:lossmap_ring}
	\end{figure*}
\fi

{\bf \noindent  Particle tracking and shower simulation -- }Dedicated simulations with SixTrack and FLUKA were performed after the test. The distribution of proton losses (i.e. inelastic events) over the IR7 collimators computed by SixTrack, {\bernhard using COLLTRACK/K2 \cite{K2,SixTrack3} routines}, was used as source term for FLUKA calculations{\bernhard ; see \cite{BrucePRSTAB}}. The latter ones incorporated a very detailed 700m long geometry model and allowed to evaluate the {\bernhard deposited energy density} in the DS magnet coils as well as the BLM signals. Results were normalised to the achieved loss rate, as measured by the FBCT.  

\begin{figure}[!t]
\centering
\includegraphics[width=0.5\textwidth]{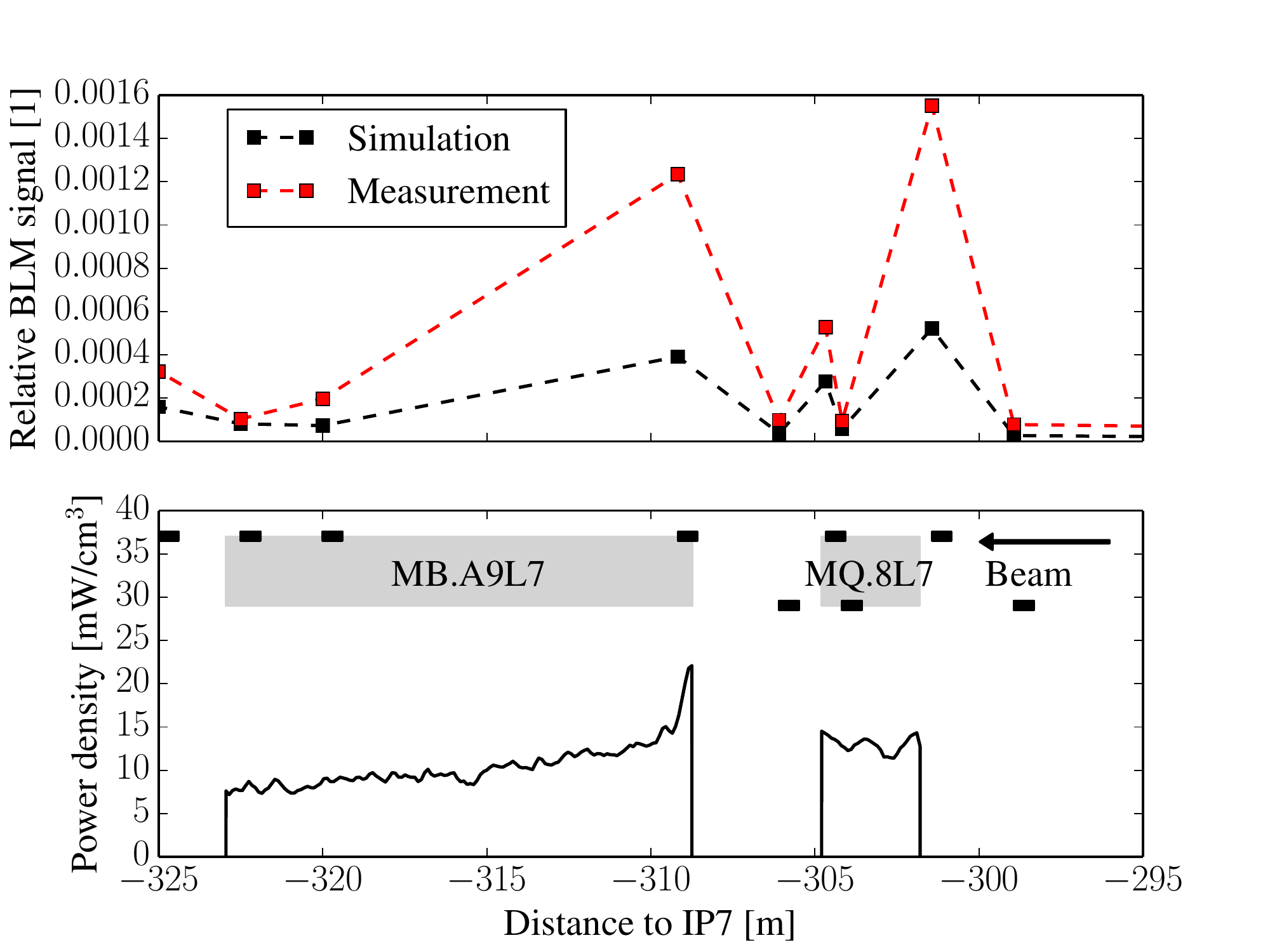}
\caption{Up: Local excerpt of the comparison of BLM signals and FLUKA simulation results for Ramp 3 of the 2013 collimation quench test, both respectively normalised to the BLM of the {\bernhard skew primary collimator (TCP.B6R7)} located at +200~m \cite{LefterisIPAC2015,AntonFLUKA}. Down: Longitudinal pattern of the power density (averaged over the inner coil radial thickness) at the transition between cells 8 and 9. 
}
\label{fig:coll:peak}\label{fig:coll:fluka_blm}
\end{figure}

\begin{figure}[!t]
\centering
\includegraphics[width=0.4\textwidth]{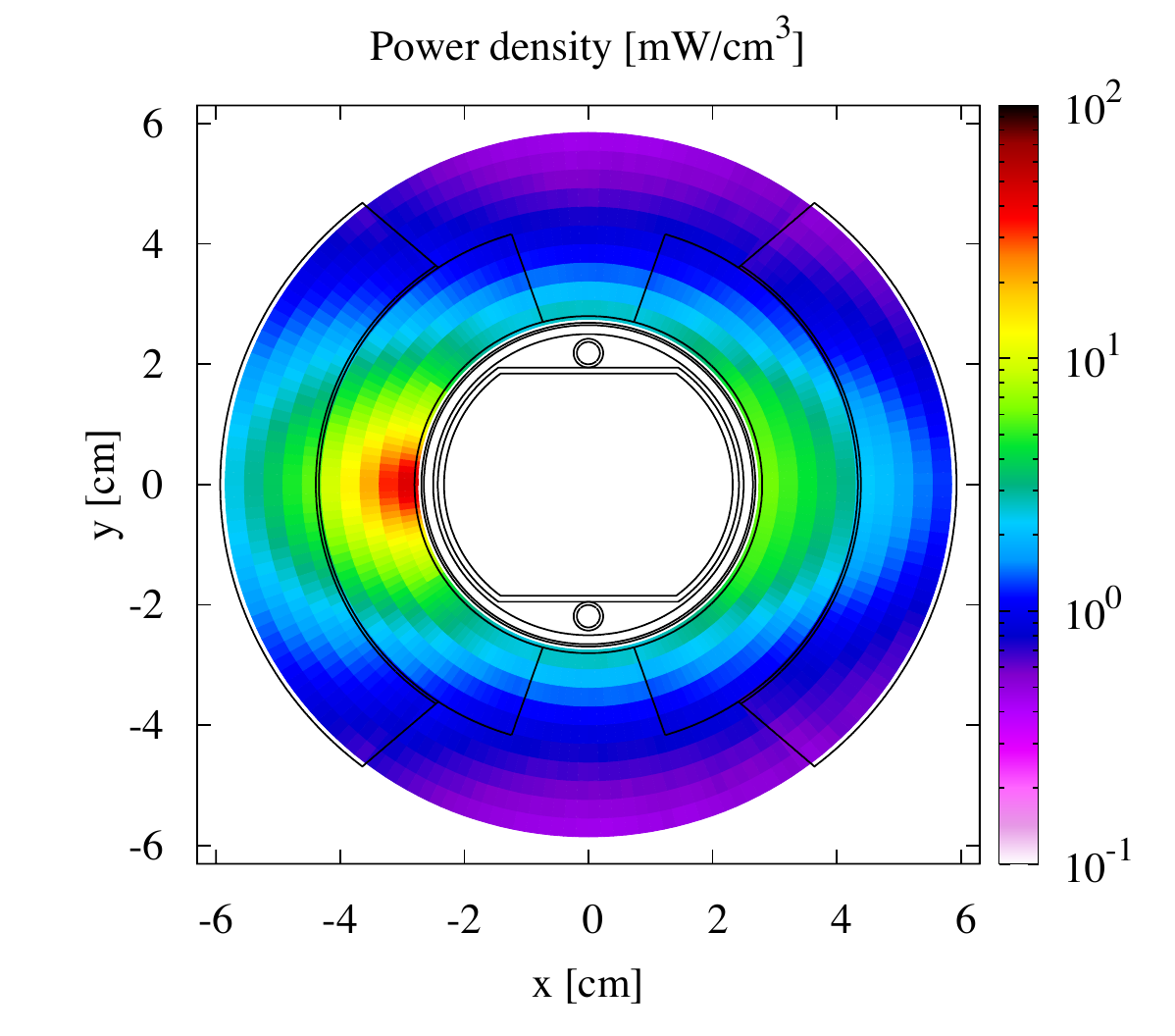}
\caption{Simulated transverse power density distribution {\bernhard from FLUKA} during the steady-state collimation quench test at the MB.A9L7 maximum in Fig.~\ref{fig:coll:peak} (down). The beam direction enters the figure. Recall from Section~\ref{Methodology} that in FLUKA coils are extruded, not adequately representing the coilends.}
\label{fig:coll:transv}
\end{figure}

In Fig.~\ref{fig:coll:fluka_blm} (down), the peak power density in the inner superconducting coils is plotted along the length of the most impacted magnets. The maximum is on the front face of the first dipole in cell 9. Figure~\ref{fig:coll:transv} shows the corresponding power density map on the magnet transverse section. The horizontal plane is mainly affected, due to the particle shower originated from protons experiencing a limited energy loss and angular kick (typically a diffractive event) in the primary collimator, leaking through the collimation system down to the DS, where they are overbent by the magnetic field towards the internal boundary of the physical aperture.


Figure~\ref{fig:coll:fluka_blm} (up) presents the comparison between measured and predicted BLM signals in the region considered here. Values are normalised to the signal of the BLM at the {\bernhard skew} primary collimator, since measurements refer to an integration time over which the loss rate was not constant. The shortest integration time of 40\,$\mu$s could instead be used for the much higher experimental signals in the collimator region, allowing there an absolute comparison confirming the full consistency of the normalisation factors adopted here \cite{AntonFLUKA}. A more exhaustive presentation of the FLUKA model and the comparison to measurements is found in \cite{LefterisIPAC2015}. Despite a remarkable agreement globally achieved over the whole IR7 insertion, calculations feature a localized underestimation of a factor of few from the end of the {\bernhard Long Straight Section up to Cell 9 (as in Fig.~\ref{fig:coll:fluka_blm}).} Such an underestimation may reasonably imply that the power density in the magnet coils was actually higher than in Figs.~\ref{fig:coll:fluka_blm} and~\ref{fig:coll:transv}.



%
{\bf\noindent Electro-thermal simulation -- } In the steady-state regime, the quench level depends on the effectiveness of the cooling to the helium bath, which was tested in dedicated experiments \cite{Granieri:2010hc}. The experiments determine the temperature rise in a stack of 10 cables under continuous heating. The stack is submerged in superfluid helium at 1.9 K and exposed to pressures of up to 100~MPa. In most magnets, the cables that are exposed to beam losses are cooled only on the inner diameter. LHC main dipole and quadrupole magnets, however, are equipped with an intra-layer spacer that is slotted in order to provide channels for the superfluid helium. In \cite{ppgranieri_steadystate}, the slots in the intra-layer spacer are assumed to be ideally effective until the strands reach the lambda temperature, i.e., the temperature when the helium in the cooling channels in the Kapton insulation stops to be superfluid. Moreover, an average pressure of 50~MPa was assumed. A more conservative model neglects the cooling to the inter-layer helium and assumes 100~MPa on the inner diameter of the heated cable. The results from both assumptions are shown in Tab.~\ref{tab:colltest:analyses}. Both models are consistent with the lower bound obtained from the collimation quench test. According to the FLUKA model, the peak of the losses was deposited in the ends of the MB magnet. As a consequence, the exact magnetic field and the cooling conditions in the position of peak losses are not accurately known; compare with a similar discussion in Sec.~\ref{sec:wire}. 
%
\begin{table}[!t]
\caption{Quench-level comparison of the FLUKA lower bound and the electro-thermal MQPD estimate in the MB.A9L7 coil for the steady-state collimation quench test. {\bernhard The upper and lower values for the electro-thermal estimate correspond to a cooling model with and without heat-flow through the intra-layer spacers, respectively.}}
\label{tab:colltest:analyses}
\centering
\begin{tabular}{cc}
\hline
 { Particle Shower}& Electro-Thermermal \\
 Calculation&Estimate\\
 \,[mW/cm$^3$]& [mW/cm$^3$]\\
\hline
$>$ 23&115-140\\
\hline
\end{tabular}
\end{table}

{\bf\noindent Discussion -- }
The collimation quench test closely reproduces an operational scenario. The relevant relationship between losses on the collimator and maximum power-deposition in the magnet coils is difficult to simulate. The large-scale FLUKA model shows overall remarkable performance \cite{LefterisIPAC2015}. In the high-loss region in the cold section of the model, however, the agreement is not satisfactory. Moreover, as the peak losses occur in the magnet ends, there are additional uncertainties due to simplifications in the geometrical model in FLUKA and the limited knowledge of cooling conditions and local magnetic field in the electro-thermal model. Only an actual quench can give more certainty in this beam-loss scenario. Nonetheless, the large-scale particle-shower model is a major step forward in our capabilities to analyze distributed events. As for the electro-thermal model, no additional insights could be gained on the efficiency of the cooling slots in the intra-layer spacers of MB and MQ magnets.  
\subsection{Dynamic Orbit-Bump Quench Test}\label{sec:dynamic}
\begin{figure}[t!]
   \centering
   \includegraphics*[width=0.5\textwidth]{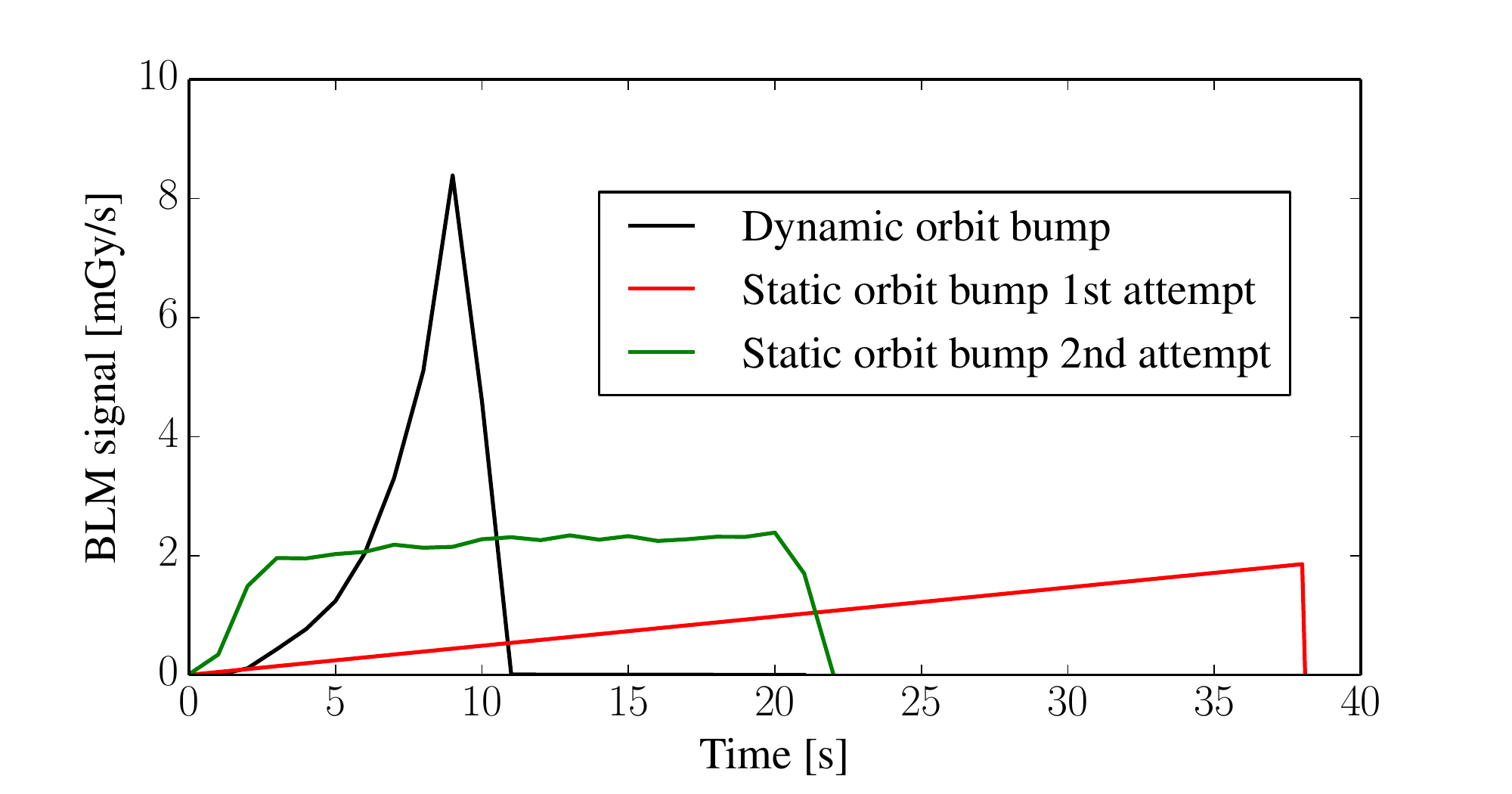}
   \caption{Comparison of the highest BLM signal time profiles for the 2010 dynamic orbit-bump quench test and the 2013 static orbit-bump quench tests. Note that in 2013 static losses were only achieved in the second attempt.
  }
   \label{thpea045-f4}
\end{figure}

{\bf \noindent Experimental setup -- }The experiment was done at 3.5 TeV beam energy. A vertical three-corrector orbit bump was formed around the main {\bernhard (horizontally defocusing)} quadrupole (MQ.14R2 at 1.9~K) and slowly
increased for $\sim$10\,s, until $\sim$60\% of the initial {\bernhard 2.54$\times$$10^{9}$}  protons were lost and the magnet quenched {\bernhard \cite{Aga_ipac11,Aga_ASC}}. 
The resulting BLM signal is shown in Fig.~\ref{thpea045-f4}.

{\bf \noindent Particle tracking -- } The spatial and time distributions of the lost particles were studied using MAD-X. A vertical orbit bump was applied around the quadrupole MQ.14R2.
In the simulations the amplitude of the orbit bump was increased by 10\,$\mu$m every 50 turns, reproducing the experimental conditions on a shorter time scale. Scaling the time axis of the normalised loss distribution to the actual loss duration provides a good qualitative fit to the normalised observed BLM signals. 

\begin{figure}[!t]
\centering
\includegraphics[width=0.5\textwidth]{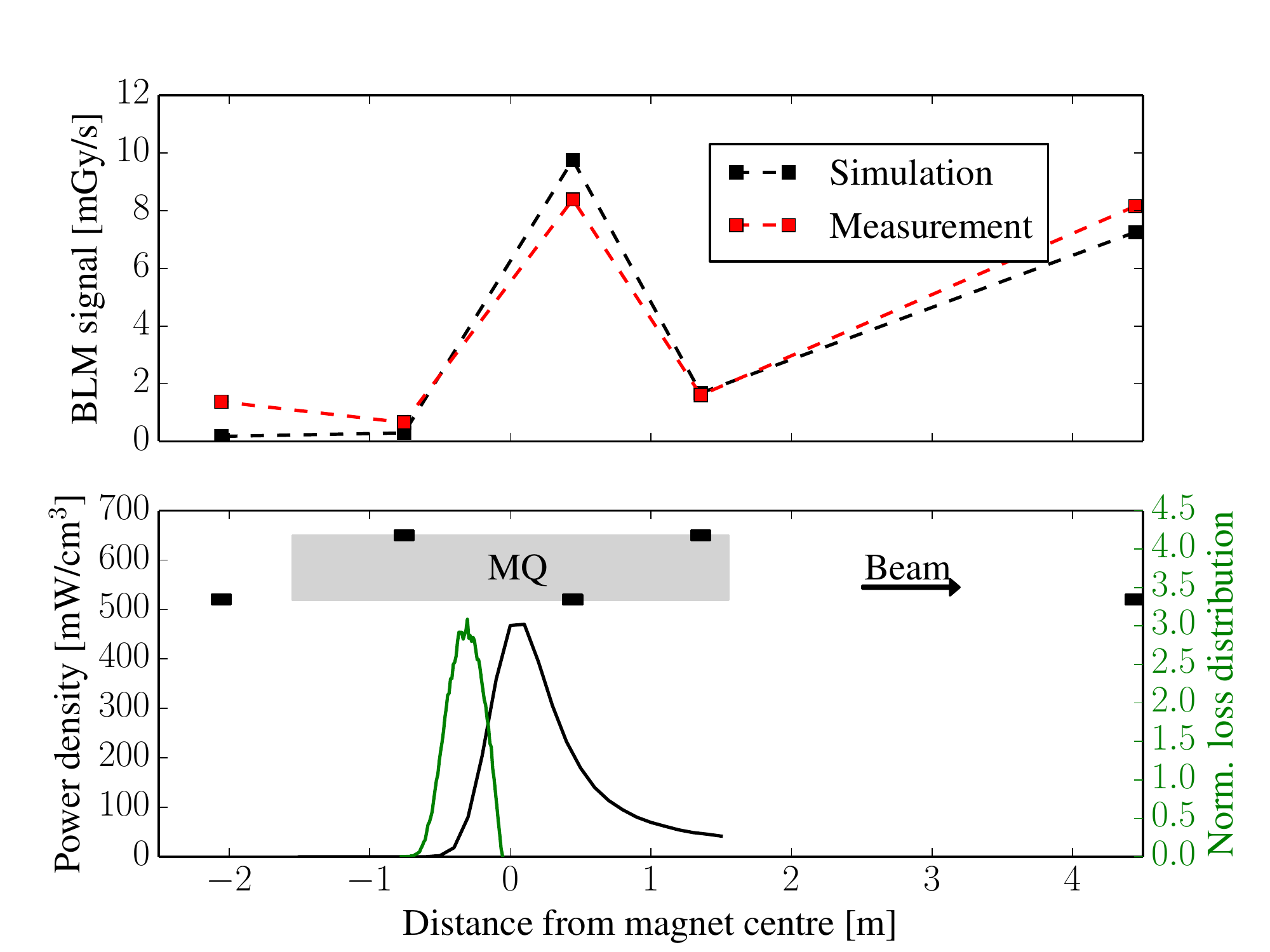}
   \caption{Up: Detail of the comparison between the BLM signal and the simulated signal from FLUKA for the dynamic orbit-bump quench test. Down: FLUKA simulated peak energy density (black) deposited in the coils and MAD-X normalised distribution of protons lost on the beam-screen (green). 
 The gray box indicates the magnet, black boxes indicate the locations of BLMs.}
   \label{fig:dyn:blms}
\end{figure}

\begin{figure}[htb]
\centering
\includegraphics*[width=0.4\textwidth]{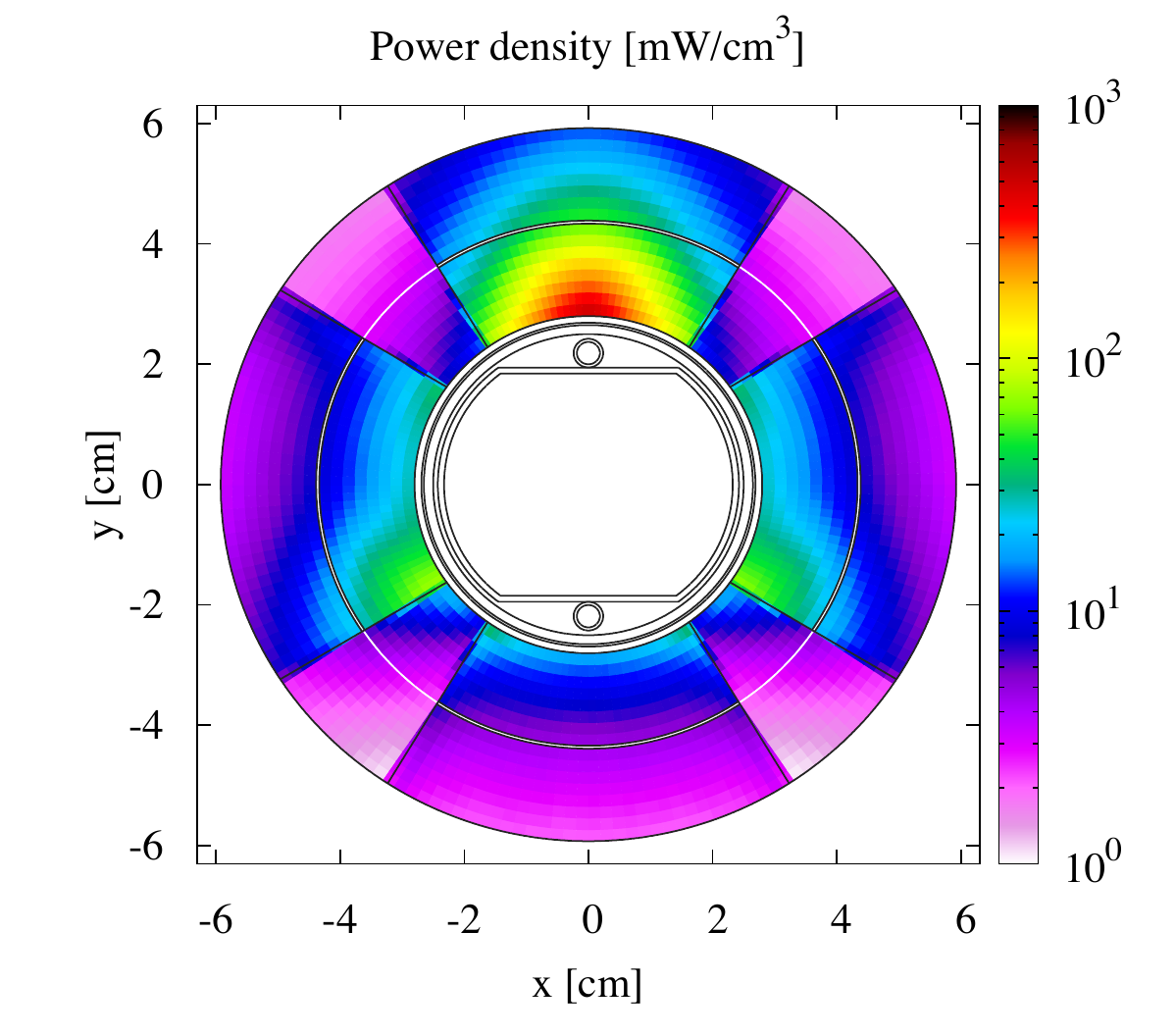}
\caption{Simulated transverse power density distribution {\bernhard from FLUKA} during the dynamic orbit-bump quench test in MQ.14R2 coils at the position where the maximum energy deposition occurs. Results correspond to 2.54$\times$10$^{9}$ protons impacting on the magnet beam screen. Spatial coordinates are with respect to the center of the vacuum chamber.}
\label{fig:dynOrbBump_ED_transverse}
\end{figure}

{\bf \noindent Particle-shower simulation -- } The loss distribution obtained from the MAD-X simulation was used as source term for the FLUKA simulation. The resulting particle shower was tracked and the energy density in the coils estimated. The number of protons lost in the last second before the dump, 2.54$\times$10$^{9}$ protons/s, was used to scale FLUKA results. {\bernhard The resulting BLM signals and energy deposition are shown in Fig.~\ref{fig:dyn:blms}. The agreement between simulation and measured BLM signals is excellent.} Figure~\ref{fig:dynOrbBump_ED_transverse} shows the simulated transverse power density distribution in the coils. The maximum power density, averaged over the affected turn, is estimated to be 208\,mW/cm$^{3}$.

{\bf \noindent Electro-thermal simulation -- }
The losses in the dynamic orbit-bump quench test were not actually steady-state. In terms of peak power at the moment of quench, the acceleration of losses increased the quench level for this loss scenario. Two alternative models were used for steady-state cooling, differing in the assumption on the effectiveness of the inter-layer cooling channels in the MQ.  {\bernhard Results} are displayed in Tab.~\ref{tab:dob:analyses}. The agreement between FLUKA and electro-thermal-model results is remarkably good. {\bernhard FLUKA values are} within the range of known uncertainty of the electro-thermal estimates.
\begin{table}[!t]
\caption{Quench-level comparison; FLUKA and the electro-thermal {\bernhard MQPD} estimate for the dynamic orbit-bump quench test. The integrated deposited energy is averaged across the cable cross-section. {\bernhard The upper and lower values for the electro-thermal estimate correspond to a cooling model with and without heat-flow through the intra-layer spacers, respectively.}}
\label{tab:dob:analyses}
\centering
\begin{tabular}{cc}
\hline
 { Particle Shower}& Electro-Thermal\\
 Calculation&Estimate\\
\,[mW/cm$^3$]& [mW/cm$^3$]\\
\hline
   208  & 180-215 \\
\hline
\end{tabular}
\end{table}

{\bf \noindent Discussion -- } 
The dynamic orbit-bump quench test was the first of its kind, producing losses over several seconds before quenching the magnet. The good agreement between measured and simulated BLM data, as in the case of the intermediate-duration orbit-bump quench test, indicates a good grasp of the beam dynamics leading up to the quench, as well as the subsequent shower development. The good agreement with the electro-thermal model, finally, makes this, one of the best understood beam-induced quenches in the LHC.

\subsection{Static Orbit-Bump Quench Test}\label{sec:slowADT}
{\bf \noindent Experimental setup -- }A local orbit bump was established around a main quadrupole magnet (MQ12.L6 at 1.9~K), such that the beam almost touched the aperture. Eight low-intensity bunches 
of 1$\times$$10^{10}$ protons each were slowly blown up using white noise excitation in the ADT. The first attempt produced linearly rising losses over 38~s and no quench {\bernhard after 6.1$\times$10$^{9}$ protons lost \cite{Agnieszka_PhDthesis}}. At the second attempt, after losing about 6.2$\times$10$^{9}$ protons at a constant rate over 20 seconds, the magnet quenched. {\bernhard The corresponding BLM signals can be seen in Fig.~\ref{thpea045-f4}}. The linear rise during the first attempt is attributed, as in the dynamic-orbit-bump test, to the Gaussian beam profile. The losses are slow enough to cut off consecutive layers of the phase-space ellipse. The rising profile, therefore, corresponds to the tails of the distribution. The same bunches were used in the second attempt. In the absence of renewed beam-profile measurements, it is assumed that the remnants of the bunches diffused into a wider and more flat distribution in between the attempts, thus, explaining the flat loss profile over time observed in the second attempt.

%
%

 
{\bf \noindent Particle tracking -- } In order to reproduce the excitation of 8 bunches in MAD-X, eight sets of simulations were performed, followed by a combined analysis. Each of the sets followed the
same procedure: first, an equilibrium beam distribution with the experimentally measured sigma was created; second, the orbital bump was established around the focusing quadrupole MQ.12L6; and finally,
the white-noise excitation with the ADT started. The sensitivity of the longitudinal distribution was tested with respect to the ADT kick strength and, as in the previous orbit-bump quench tests, to aperture restrictions. {\bernhard Additional} studies have shown that 
increase of the kick strength leads to a decrease in the height of the distribution, a shortening of the loss duration, and a longer longitudinal distribution. A weak value of ADT kick strength within the realistic parameter range produces the most realistic results. The impact angle depends only on the integral magnetic field of the quadrupole, seen by the particles. 

\begin{figure}[!t]
\centering
\includegraphics[width=0.5\textwidth]{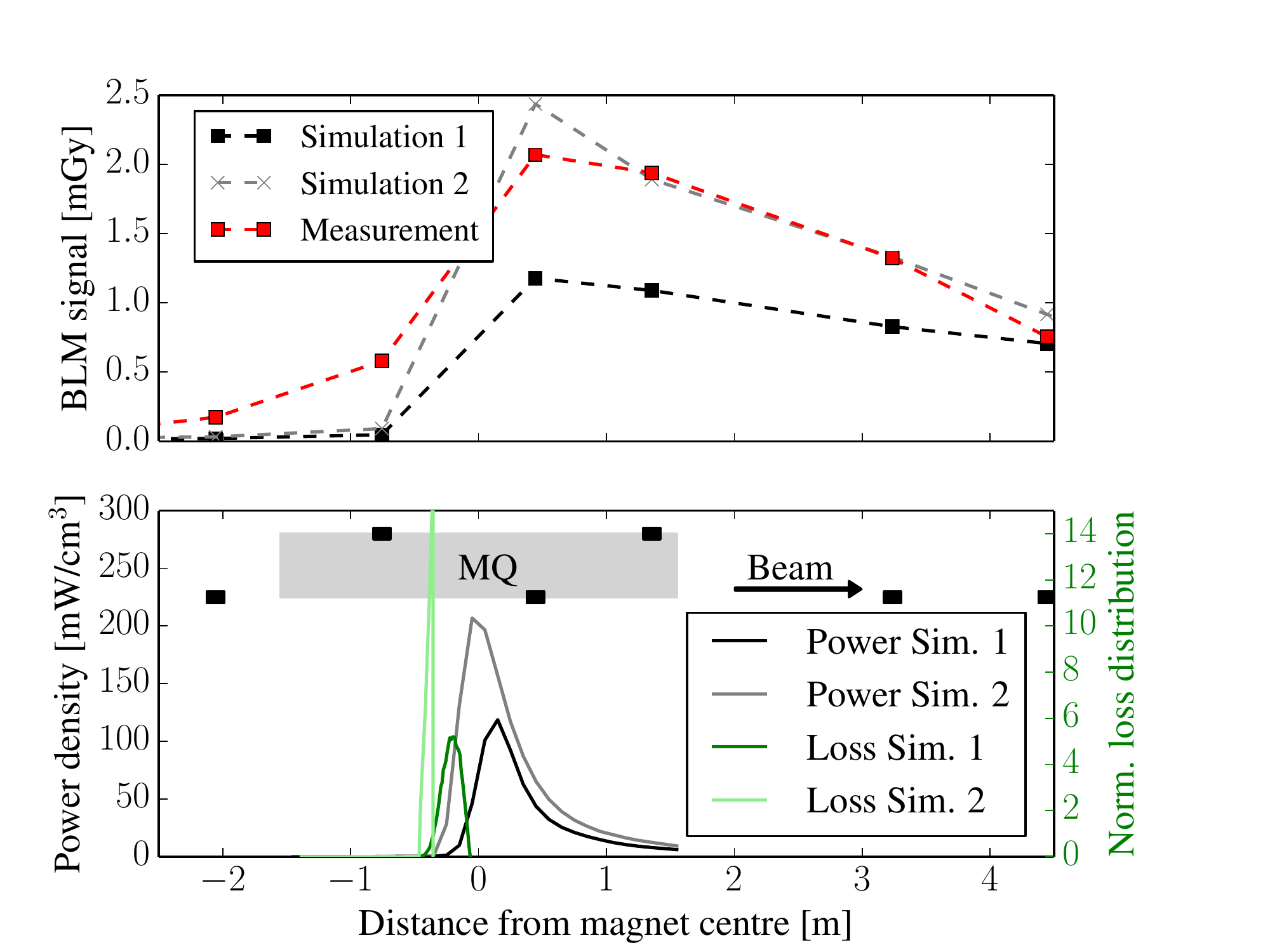}
   \caption{Results of particle-tracking and particle-shower simulations. Simulation 1 considers a smooth beam screen, whereas Simulation 2 considers a 20-cm-long, 30-$\mu$m-high aperture restriction close to the magnet center. 
   Up: Detail of the comparison between the BLM signal and the simulated signal from FLUKA for the static orbit-bump quench test. Down: FLUKA simulated peak energy density (black) deposited in the coils and MAD-X normalised distribution of protons lost on the beam-screen (green). 
 The gray box indicates the magnet, black boxes indicate the locations of BLMs. }
   \label{fig:ssqt:blms}
\end{figure}
\begin{figure}[t]
\centering
\includegraphics[width=0.4\textwidth]{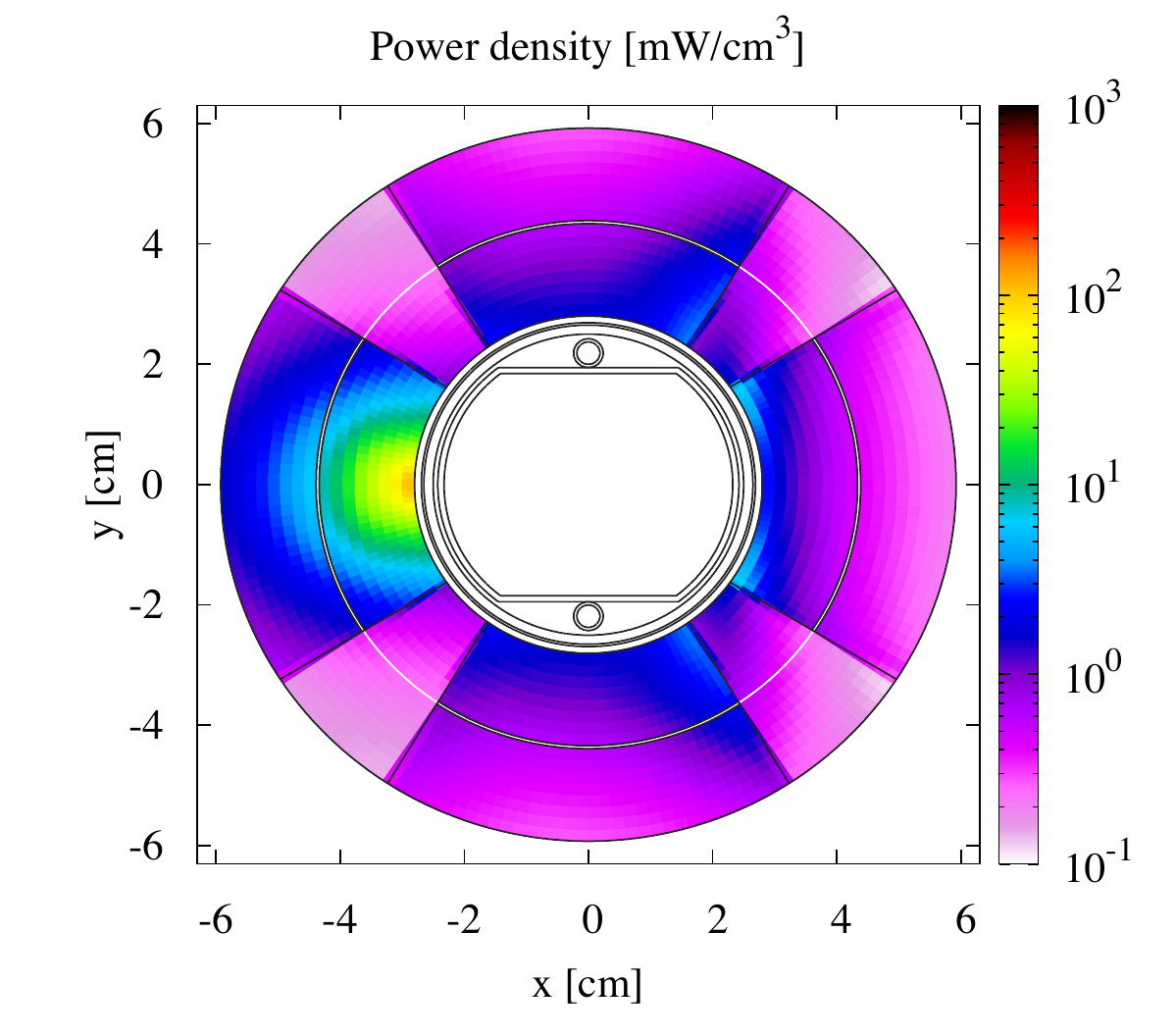}
\caption{Simulated transverse power density distribution during the static orbit-bump quench test in MQ.12L6 coils at the location where the maximum energy deposition occurs. Results correspond to 3$\times$10$^{8}$ protons impacting on the magnet beam screen. 
}
\label{fig:adtslow_smooth-surface:transv}
\end{figure}

In the subsequent particle-shower simulation moderate agreement of measured and simulated BLM signals was found; {\bernhard see Fig.~\ref{fig:ssqt:blms}}. This came as a surprise, after the successful reproduction of BLM signals during the intermediate-duration orbit-bump quench test and the dynamic orbit-bump quench test. A study of the impact of small aperture restrictions was carried out to check whether a geometric effect of this kind could explain the discrepancy. A 20-cm-long and 30-$\mu$m-high step into the otherwise smooth aperture was found to have a strong influence on the overall loss distribution of particles. Since the impact angle of the particles is small, losses peak on the onset of the restriction, leaving the surface behind it in its shadow. Results can be seen in Fig.~\ref{fig:ssqt:blms}, where Simulation 1 refers to a simulation with smooth beam screen, and Simulation 2 to a simulation with a localized aperture restriction. Since the impact angle of particles varies roughly linearly along the length of the quadrupole magnet, the shifting of the distribution implies an increase by 100\% of the average impact angle of protons on the beam screen.  


{\bf \noindent Particle-shower simulation -- } In the static orbit-bump quench {\bernhard test the magnet quenched} from a near constant particle loss rate of about 3$\times$10$^{8}$ protons/s. Fig.~\ref{fig:adtslow_smooth-surface:transv} shows the transverse profile of power density at maximum along the longitudinal axis for smooth surface. Since the loss distribution is similar to that of intermediate-duration orbit-bump quench test, the power density profile is also similar. The maximum power density occurs in the internal coil of the magnet and is about 41\,mW/cm$^3$, averaged over the cable cross-section. The comparison of predicted and measured BLM signals, as well as the corresponding power-deposition in the coil, is shown in Fig.~\ref{fig:ssqt:blms}. The second simulation, assuming a localized aperture restriction, yielded an 80\% higher maximum power density mainly due to the increase in impact angle of the lost protons.

{\bf \noindent Electro-thermal simulation -- }
Two alternative models were used for steady-state cooling, differing in the assumption on the effectiveness of the inter-layer cooling channels in the MQ. Results are displayed in Tab.~\ref{tab:ss:analyses}. {\bernhard Note that the first attempt has not been simulated by MAD-X and FLUKA. The displayed values are rescaled by use of the BLM signals in Fig.~\ref{thpea045-f4}. We estimate the uncertainty introduced by this scaling to be 10\%. The more conservative assumptions give the better agreement with the FLUKA analysis and are, therefore, used as the baseline. 

As} the test featured the longest duration of continuous losses of all studied events, another explanation of the low quench level has been suggested: that the quench level is determined not by the heat transport through the cable insulation, but by the heat transport towards the heat exchanger, thus, requiring the modeling of an entire magnet cross-section, rather than just a turn in the coil. Further studies in this direction will be carried out, in continuation of previous work \cite{bocianSS, brucebocian}.

\begin{table}[!t]
\caption{Quench-level comparison; FLUKA and the electro-thermal MQPD estimate for the static orbit-bump quench test. Simulation 1 and 2 correspond to MAD-X models without and with an aperture restriction, respectively; compare with Fig.~\ref{fig:ssqt:blms}. {\bernhard The upper and lower values for the electro-thermal estimate correspond to a cooling model with and without heat-flow through the intra-layer spacers, respectively.}}
\label{tab:ss:analyses}
\centering
\begin{tabular}{cccc}
\hline
 Attempt&Simulation&{ P. Show.}& El.-Therm.\\
& &[mW/cm$^3$]& [mW/cm$^3$]\\
\hline
1st &1&  $>${\bernhard 36}  & 80-100 \\
1st &2&  $>${\bernhard 61}  & 80-100\\\
2nd&1&{\bernhard 43}&70-88\\
2nd&2&{\bernhard 72}&70-88\\
\hline
\end{tabular}
\end{table}

{\bf \noindent Discussion -- }
The slow blow-up of the beam, generated with the ADT transverse damper, was effective in generating near steady-state losses over 20 seconds, strong enough to eventually quench an MQ magnet. The test represents, therefore, a good benchmark for the steady-state electrothermal models. The sizable discrepancy between measured and simulated BLM signals led to a comprehensive parametric study in the particle-tracking simulations. An aperture restriction {\bernhard on the scale of several tens of micrometers} could best account for the discrepancy. To confirm or rule out this explanation, a repetition of the test would have to be carried out in a different location.

\section{Conclusions and Outlook}\label{Conclusion}
\subsection{Summary of Quench Tests}

The large-kick event and the short-duration collimation quench test are interesting benchmarks for the particle-tracking and particle-shower simulations, because the electro-thermal MQED estimate is not expected to have a sizable error as a consequence of the negligible contribution of heat transfer processes. The large-kick event indeed confirmed the MQED estimate. The collimation quench test led to a 50\% overestimation of the MQED estimate by the FLUKA result. This is likely due an inaccuracy in the FLUKA geometrical model. Without meaningful BLM signals (saturated channels), however, there is no clear indication where to search for such a discrepancy. The 50\% error must, thus, be regarded as a measure of the error that may affect other FLUKA analyses in the absence of validation data. 

In the intermediate-duration regime we note that a precise timing and an adequate time resolution of signals is of paramount importance. Future tests in this regime should include synchronous measurement of BLM, QPS, and FBCT signals. For a better understanding of intermediate-duration losses due to dust particles, a test should create milli-second losses with a smooth time distribution in an MB or MQ magnet (at 1.9 K). The orbit-bump quench test featured losses that peaked of several tens of microseconds every three to four turns of the excited bunch. It is suspected that this substructure of short loss spikes has led to the surprisingly high quench level, four times above the expected one.

The testing of quench levels in the steady-state regime is of importance, for example, for the strategies for future collimation upgrades. In this sense the empirical result of the collimation quench test (no quench for 5.8~MJ on the primary collimator within \hbox{15 s}) gives actionable information and the test will be repeated at higher beam energies for protons and ions. Based on the dynamic and static orbit-bump quench tests, we conclude that the semi-empirical steady-state cooling model seems to suggest that inter-layer spacers are not having a large effect on the steady-state quench level. An overview of the analysis results is given in Tab.~\ref{tab:summary}.

\begin{table*}[t]
\caption{Overview of the presented analyses. BLM validation indicates the level of agreement between particle-tracking and particle-shower simulations with BLM data. Quench level consistency indicates the agreement between quench-level data obtained from particle-shower simulation results with the electro-thermal estimates. `Good' indicates agreement within 20-30\%, `average' around 50\%, and `poor' larger than 100\%}
\label{tab:summary}
\centering
\begin{tabular}{ccccc|c|c|l}
\hline
Regime&Method&Type&Temp.&  $I/I_{\rm nom}$&BLM &Quench Level &Sources of Uncertainty \\
&&&[K]&[\%] & Validation&Consistency& \\\hline\hline
short&kick&MB&1.9&6&good&good&\begin{tabular}{l}Tracking uncertainty.\end{tabular}\bigstrut\\\hline
short&collimation&MQM&4.5&46/58&n/a&average&\begin{tabular}{l}Saturated BLM signals. No FLUKA validation.\end{tabular}\bigstrut\\\hline
intermediate&wire scanner &MBRB&4.5&50&good&average&\begin{tabular}{l}Timing uncertainty. Quench in magnet end.\end{tabular}\bigstrut\\\hline
intermediate&wire scanner &MQY&4.5&50&good& consistent&\begin{tabular}{l}No quench.\end{tabular}\bigstrut\\\hline
intermediate&orbit bump&MQ&1.9&54&good&poor&\begin{tabular}{l}Timing uncertainty. Inaccurate heat-transfer model.\end{tabular}\bigstrut\\\hline
steady-state&collimation&MB&1.9&57&poor&consistent&\begin{tabular}{l}No quench. Peak loss in magnet end. \\
Moderate FLUKA agreement with BLM signals.\\ Intra-layer spacer cooling efficiency. \end{tabular}\bigstrut\\\hline
steady-state&static orbit bump&MQ&1.9&54&average&average&\begin{tabular}{l}Sensitivity to surface roughness. \\Intra-layer spacer cooling efficiency. \end{tabular}\bigstrut\\\hline
steady-state&dyn. orbit bump&MQ&1.9&0.47&good&good&\begin{tabular}{l}Intra-layer spacer cooling efficiency. \end{tabular}\bigstrut\\
\hline
%
%
%
\end{tabular}
\end{table*}
\subsection{Impact on Quench Level Estimates}
For single-turn losses, i.e. losses of nanosecond duration,  we have learnt to trust the electro-thermal model, which is based on the strand enthalpy margin. Note, however, that for the operation of the LHC, the shortest duration that is resolved by the BLM system is 40~$\mu$s. Based on the intermediate-duration orbit-bump quench test we should revise our quench-level estimate upward by a factor four in the millisecond time range. The analysis of the quench test has given room for speculation that the high observed quench level may be due to the temporal substructure of the beam loss. If losses of microsecond duration are cooled much more efficiently than slower losses, this would be an indication that also the quench-level estimates in the microsecond range need to be revised upwards. Moreover, it is not clear how this uncertainty scales up to higher energies. We use the same factor four at 7~TeV as for 4~TeV. 

As for steady-state losses, the {\bernhard testing of the semi-empirical model used in the electro-thermal estimates} is not quite conclusive. Whereas the dynamic orbit-bump test seems to indicate a larger quench level, the static orbit-bump test points to lower levels. This could be due to a low efficiency of the intra-layer spacer's cooling channels, or due to different bottleneck in the heat transfer to the heat exchanger tube. Additional numerical and experimental studies are required. Moreover, it must be noted that the semi-empirical model is based on measurements on cable stacks with the insulation scheme of LHC main magnets. Quadrupoles and separation dipoles in the dispersion suppressor, matching section, and inner-triplet region are equipped with different insulation schemes. Clearly, a direct use of the present model for those magnet types is doubtful.

\begin{figure}[t!]
   \centering
   \includegraphics*[width=0.5\textwidth]{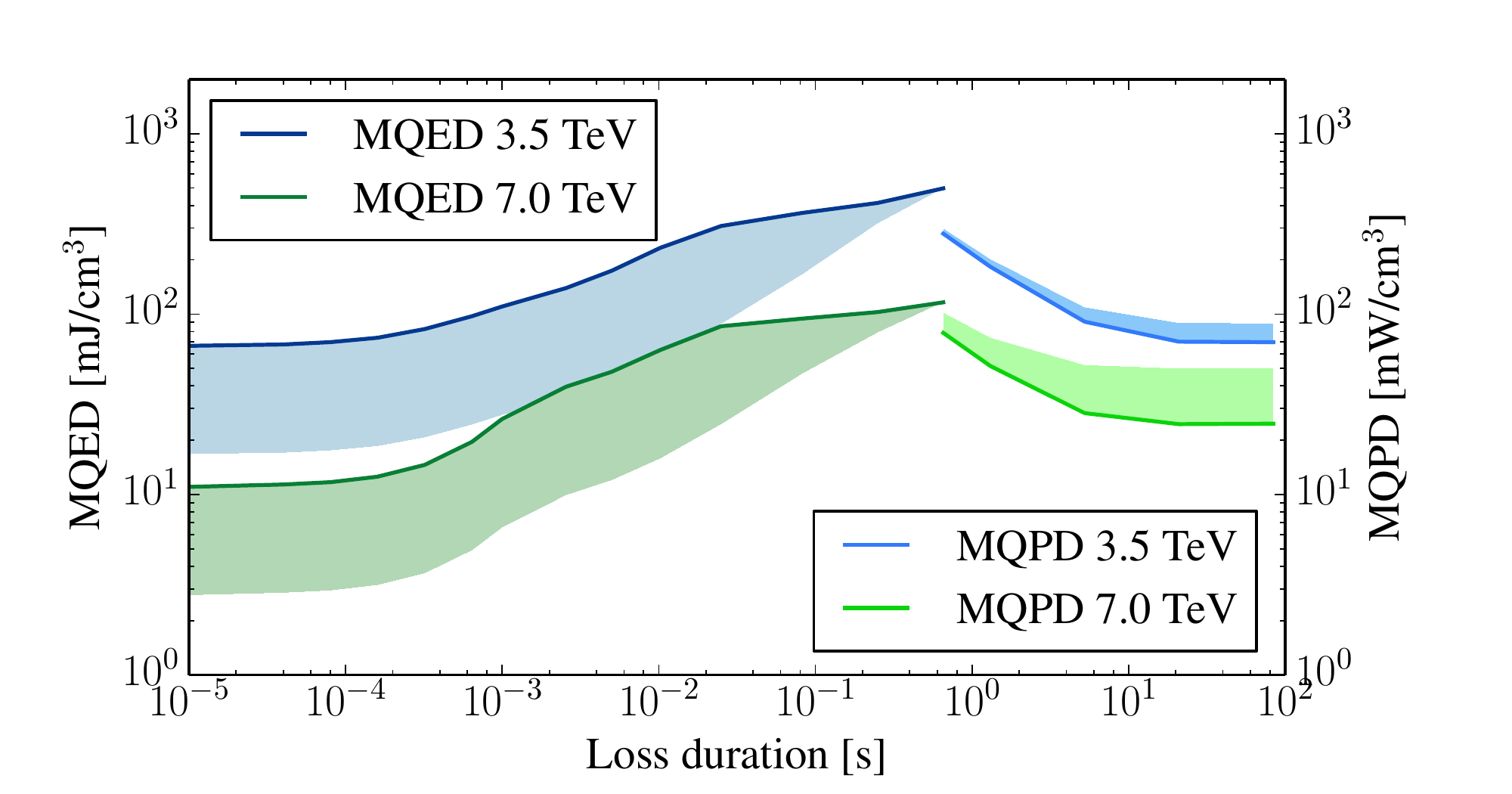}
   \caption{Electro-thermal quench level estimates for the inner layer of the LHC main bending magnet. Shading in the MQED estimates indicates uncertainties following from the analysis of the intermediate-duration orbit-bump quench test. Shading in the MQPD estimates indicates uncertainties due to the unknown cooling efficiency of intra-layer spacers in the main magnets of the LHC.}
   \label{fig:final}
\end{figure}

Figure~\ref{fig:final} provides a summary of the lessons learnt for quench levels. The shaded areas represent the uncertainty ranges of our numerical models at 3.5 and 7~TeV for rectangular loss pulses of durations in the intermediate-duration and steady-state regimes. The solid lines indicate the new baseline for the setting of BLM thresholds. For intermediate-duration losses, the more more progressive branch was selected, in line with the intermediate-duration dynamic orbit bump quench test, was retained, whereas for losses in the steady-state regime, the more conservative branch was chosen, in line with the static orbit-bump quench test. For magnet types that have not been tested with steady-state quench tests, such as the matching-section quadrupoles, the separation dipoles, and the triplet quadrupoles, an even more conservative model was retained, representing the cable insulation as a solid layer of Kapton, thus, neglecting any aid to heat transport through the insulation, by the superfluid helium.

\subsection{Future Tests}
Quench tests and beam-loss events during Run 1 have substantially improved our understanding of various beam-loss scenarios, and of the quench levels of main magnets in the LHC. We have shown that, based on good knowledge of the initial conditions and on validation data, the numerical models reproduce beam-loss events to a remarkable degree. Based on this new-found confidence, quench tests can be considered one of the most accurate means to test and validate electro-thermal estimates of quench levels under realistic conditions. 

For single-turn losses, which are relevant, for example, in case of asynchronous beam dumps, another collimator quench-test analogous to the test on the Q6 magnet described in Section~\ref{sec:Q6} is under preparation. To improve the model validation, beam-loss monitors with higher sensitivity and dynamic range will be used. 

In the intermediate-duration regime, {\bernhard during} Run 2 of the LHC a significant number of beam-losses due to collisions of the proton beam with dust particles may provoke beam-induced quenches. Based on the knowledge of beam parameters and particle-shower models, the corresponding BLM data will allow to obtain additional information of quench levels in the relevant time range. To prepare for the need of a controlled quench test with diagnostic equipment (oscilloscope, etc.), we are studying the possibility of causing losses in the millisecond regime through a combination of a local orbit bump in a main quadrupole and a fast current decay in a warm dipole leading to an orbit distortion \cite{tobias_thesis}. 

Finally, for steady-state losses we advocate the repetition of the static orbit-bump quench test in a different location in the arc to better understand the discrepancies observed in analysis of the first test of this type. A repetition could confirm or rule out the presence of a small geometrical obstruction that may have influenced the outcomes of the first test. Moreover we suggest to repeat the test in standalone quadrupoles with different insulation schemes, such as MQM, MQY, MQXA, and MQXB types. In the absence of additional sub-scale experiments, these tests could provide information on the steady-state cooling efficiency in those magnet types.   

{\bernhard In 2013 there was not the opportunity to execute the steady-state collimation quench tests with ions. 
A dedicated test in Run 2 will allow to directly determine the quench limit for this loss-scenario. The same approach should be followed for a proton collimation quench test at increased beam energy; the result of this tests provides direct input for the optimized setting of BLM thresholds, the minimum allowed beam lifetime, and on the collimator upgrade for High Luminosity LHC.}

\subsection{Conclusion}
The preparation and organisation of quench tests and the analysis of beam-loss events have been highly collaborative and multi-disciplinary efforts, stretching over the past several years. We have found that, with good knowledge of initial conditions and sufficient data for validation, particle-tracking and particle-shower simulations provide, in the best cases, results in 20\% agreement with BLM signals in the region of peak losses. This level of accuracy allows to make quantitative statements on the range of validity of electro-thermal quench-level estimates. The gained knowledge on beam-loss scenarios and quench levels is currently being applied in the setting of BLM thresholds in the LHC for Run 2.

\end{document}